\documentclass[11pt, a4paper]{article}
\usepackage{jheparxiv}
\usepackage[utf8]{inputenc}
\usepackage{amsmath}
\usepackage{amsfonts}
\usepackage{amssymb}
\usepackage{amsthm}
\usepackage{latexsym}
\usepackage{mathrsfs}
\usepackage{braket}	%Bra Ket Notation
\usepackage{mathtools}
\usepackage{graphicx}
\usepackage{color}
\usepackage{xcolor}
\usepackage{slashed}
\usepackage{twistor}
\usepackage[all]{xy}
\usepackage{tikz-cd}

\usepackage{cancel}

\usepackage{todonotes}
\reversemarginpar %%%% added so that comments and equations labels don't overlap

\newcommand{\ud}{\mathrm{d}}
\renewcommand{\Q}{q} %% this is my command for the $p'-p$. Has now been set of q.
\newcommand{\q}{\ell} %% this is my command for TOTAL momentum transfer $k+p'-p = k+q$. Has now been set to \ell.
\theoremstyle{definition}
\newtheorem{definition}{Definition}
\title{Scattering amplitudes for self-force}
\author[a]{Tim Adamo,}
\author[a]{Andrea Cristofoli,}
\author[b]{Anton Ilderton}
\author[a]{\& Sonja Klisch}

\affiliation[a]{School of Mathematics and Maxwell Institute for Mathematical Sciences \\
        University of Edinburgh, EH9 3FD, United Kingdom}
        
\affiliation[b]{Higgs Centre, School of Physics \& Astronomy \\
		University of Edinburgh, EH9 3FD, United Kingdom}

\emailAdd{t.adamo@ed.ac.uk}
\emailAdd{acristof@ed.ac.uk}
\emailAdd{anton.ilderton@ed.ac.uk}
\emailAdd{s.klisch@ed.ac.uk}

\abstract{The self-force expansion allows the study of deviations from geodesic motion due to the emission of radiation and its consequent back-reaction. We investigate this scheme within the on-shell framework of semiclassical scattering amplitudes for particles emitting photons or gravitons on a static, spherically symmetric background. We first present the exact scalar 2-point amplitudes for Coulomb and Schwarzschild, from which one can extract classical observables such as the change in momentum due to geodesic motion. We then present, for the first time, the 3-point semiclassical amplitudes for a scalar emitting a photon in Coulomb and a graviton on linearised Schwarzschild, outlining how the latter calculation can be generalized to the fully non-linear Schwarzschild metric. Our results are proper resummations of perturbative amplitudes in vacuum but, notably, are expressed in terms of Hamilton's principal function for the backgrounds, rather than the radial action.}

\begin{document}

\maketitle
  
\section{Introduction}

Perturbation theory in general relativity has been developed extensively over the past 100 years using the two-body problem as a natural laboratory (cf., \cite{Maggiore:2007ulw,Blanchet:2013haa,poisson2014gravity,Maggiore:2018sht}). In this case, the presence of multiple scales allows for several perturbative schemes which may be investigated individually, as in the effective field theory approach to the two-body problem first pioneered by Goldberger and Rothstein~\cite{Goldberger:2004jt}; see~\cite{Goldberger:2007hy,Rothstein:2014sra,Porto:2016pyg} for reviews of subsequent developments. Among these schemes, the post-Minkowskian (`PM') expansion~\cite{Bertotti:1956pxu,Bertotti:1960wuq,Rosenblum:1978zr,Westpfahl:1979gu,Portilla:1979xx,Portilla:1980uz,Bel:1981be,Damour:1981bh, Westpfahl:1985tsl}, valid for weak gravitational fields but with no restrictions on velocity, has received renewed attention. This interest follows a remarkable state-of-the-art calculation, building upon~\cite{Neill:2013wsa,Cheung:2018wkq}, for the PM expansion of the conservative potential of a compact binary system, using on-shell scattering amplitudes~\cite{Bern:2019nnu,Bern:2019crd}. This calculation and subsequent works~\cite{Bern:2020gjj,DiVecchia:2021ndb,Bern:2021dqo,DiVecchia:2021bdo,Bjerrum-Bohr:2021din,Damgaard:2021ipf,Brandhuber:2021eyq,Bern:2021yeh,Manohar:2022dea,DiVecchia:2022piu} have demonstrated the possibility of systematically organizing the PM expansion solely in terms of on-shell amplitudes in Minkowski spacetime and their classical limits, bypassing ordinary perturbative methods in classical relativity and providing an alternative way to understand the two-body problem (cf., \cite{Iwasaki:1971iy,Holstein:2004dn,Kosower:2018adc} and reviews~\cite{Bjerrum-Bohr:2022blt,Kosower:2022yvp,Buonanno:2022pgc}).

In particular, it has been demonstrated -- for instance, using the formalism developed by~\cite{Kosower:2018adc,Maybee:2019jus,delaCruz:2020bbn,Cristofoli:2021vyo,Cristofoli:2021jas} -- that the PM expansion of classical observables such as scattering angle, waveform and power emitted are determined by corresponding on-shell, perturbative scattering amplitudes. In essence, the classical observable, to a given PM precision, is determined by an on-shell phase space integral over the classical limit of on-shell scattering amplitudes (or products thereof) computed in the ordinary perturbative expansion of a quantum field theory. For example, the classical waveform for gravitational radiation emitted by the scattering of two Schwarzschild black holes is determined to leading PM order by the tree-level 5-point amplitude for two massive scalars to scatter and emit a single graviton in the field theory of massive scalars minimally coupled to general relativity~\cite{Cristofoli:2021vyo}.

Consequently, it is natural to consider whether the same tools can be applied to other perturbative expansions relevant to the two-body problem in general relativity. One example is the self-force expansion~\cite{Cutler:1994pb,Gralla:2008fg,Barack:2009ux,Gralla:2009md,Poisson:2011nh,Barack:2018yvs}, whose relation to on-shell amplitudes has been the subject of recent papers~\cite{Adamo:2022qci,Barack:2023oqp}. This expansion assumes the existence of an exact solution to the Einstein field equation, referred to as the \emph{background} and it is defined by studying deviations from geodesic motion in powers of a dimensionless parameter given by the mass of the particle and a natural mass scale associated with the background.

The self-force expansion on a Schwarzschild background is perhaps the most notable example, due to its significance in the effective-one-body description which accurately describes the two-body problem in general relativity for a (non-spinning) binary system of masses $m$ and $M\gg m$~\cite{Damour:2009sm,Akcay:2012ea,Nagar:2022fep,Albertini:2022dmc}.
At zeroth-order in the mass ratio $m/M$, the motion of the probe particle follows a geodesic, and no radiation is emitted. At next-to-leading order, the emission of gravitational waves due to the particle's acceleration and its consequent backreaction is taken into account by solving Einstein's field equations to the same order~\cite{Detweiler:2008ft,Blanchet:2009sd,Hopper:2015icj,Wardell:2021fyy,Lynch:2023gpu}, or equivalently, considering EFT in curved spacetime~\cite{Galley:2006gs,Galley:2008ih}. Interestingly, most of the literature on this topic has focused on the study of self-force for initially \emph{bound} orbits, while the investigation of the \emph{unbound} case has only gained attention quite recently~\cite{Hopper:2017qus,Hopper:2017iyq,Gralla:2021eoi,Long:2021ufh,Long:2022sdq,Barack:2022pde,Barack:2023oqp,Whittall:2023xjp}. 
This setting seems highly amenable to scattering amplitude methods, where initial data is expressed in terms of on-shell, unbound states. 

Indeed, the self-force expansion on a gravitational plane wave background (where all orbits are unbound and wavefunctions for external scattering states can be determined exactly~\cite{Ward:1987ws,Adamo:2017nia,Adamo:2020qru}) has recently been pursued using precisely these methods. Classical on-shell observables, such as impulse and waveform, can be derived directly from on-shell scattering amplitudes on the plane wave background~\cite{Adamo:2022rmp,Adamo:2022qci,Cristofoli:2022phh}. Further, the Penrose limit argument~\cite{Penrose:1976} suggests that the self-force expansion on a gravitational plane wave serves as a useful toy model for understanding self-force on \emph{any} background, along a null geodesic.

Building on the findings of~\cite{Adamo:2022qci}, an intriguing question arises: if the classical limit of on-shell amplitudes on a plane wave background encodes the self-force expansion for unbound particles, what are the necessary building blocks to compute self-force corrections to scattering observables on a Schwarzschild background?
In this paper, we provide a straightforward answer: the required building blocks are represented by semiclassical on-shell amplitudes on the corresponding Schwarzschild background, constructed within the perturbiner approach (cf., \cite{DeWitt:1967ub,Arefeva:1974jv,Abbott:1983zw,Jevicki:1987ax,Rosly:1996cp,Rosly:1996vr,Rosly:1997ap,Selivanov:1999as,Mizera:2018jbh,Garozzo:2018uzj,Cho:2021nim,Lee:2022aiu,Cho:2022faq}). Here tree-level scattering amplitudes are computed from solutions of the classical equations of motion, determined by the multiplicity of the scattering process of interest, and by the quantum numbers of the scattered states. While this is also the natural input for standard (i.e.,~Feynman diagram) calculations of scattering amplitudes, the power of the perturbiner approach is that it can be used even when the S-matrix does not exist, as on a Schwarzschild background~\cite{DeWitt:1953ipa,Hawking:1974rv,Hawking:1975vcx,Gibbons:1975kk,Gibbons:1975jb,Woodhouse:1976fe}. We show that the building blocks of self-force corrections constructed in this way are controlled by Hamilton's principal function on the background. Furthermore, we demonstrate that they may also be interpreted in terms of resummed perturbative amplitudes. We thus argue that our approach provides a conceptual pathway to the self-force approximation for unbound orbits solely in terms of on-shell amplitudes in vacuum\footnote{At the geodesic level, the possibility of exploring all-order results from resummed scattering amplitudes in vacuum has been investigated in~\cite{Cheung:2020gbf}, using an algebraic relation between scattering amplitudes and the Hamiltonian in an isotropic gauge~\cite{Damour:2017zjx,Bern:2019crd,Kalin:2019rwq,Bjerrum-Bohr:2019kec}, valid to all orders only in 4-dimensions~\cite{Cristofoli:2020uzm}.}.

Our focus is on the scattering of scalar particles and the leading order emission of radiation in static spherically symmetric backgrounds. We find it useful to consider  electromagnetism alongside the more complicated gravitational setting (see~\cite{Bern:2021xze,Bern:2023ccb} for studies of the two-body probem in scalar QED using the electromagnetic analogy of the PM expansion), so our investigation concerns scattering in Coulomb and Schwarzschild backgrounds, respectively.

\medskip

This paper is organized as follows: In Section~\ref{sec:intro-to-semi} we review the perturbiner method and introduce the notion of \emph{semiclassical scattering amplitudes}, the calculation of which will be the focus on much of this paper. Section~\ref{sec:2pt} applies this formalism to elastic scattering of massive scalars on Coulomb and Schwarzschild backgrounds, where we introduce a novel method to define the semiclassical wavefunctions in terms of Hamilton's principal function (HPF) for the background and certain `matching coefficients' to ensure proper asymptotic behaviour. We see that this reproduces the well-known expressions for elastic scattering and geodesic motion in terms of the radial action of the background.  

We then proceed to the computation of the semiclassical photon emission amplitude in Section~\ref{sec:photons}. We see that this is is controlled by the HPF (rather than the radial action), and demonstrate that in the classical, weak field limit the semiclassical amplitude on the Coulomb background gives the probe limit of the classical 5-point photon emission amplitude from two scalars. This implies that the classical, probe limit of the 5-point amplitude is in fact a linear function of the HPF itself, highlighting that it is the HPF, rather than radial action, which controls radiation. Section~\ref{sec:gravitons} deals with the semiclassical graviton emission amplitude, where the definition of the emitted graviton state presents new complications. While we are only able to define the amplitude schematically on the full Schwarzschild metric, linearising the background enables a more explicit but still rich computation. The semiclassical amplitude is again controlled by the HPF, with its classical weak field limit reproducing the classical probe limit of the 5-point graviton emission amplitude from massive scalars in Minkowski spacetime. Section~\ref{sec:concl} concludes with a discussion of future directions and how classical, self-force observables can be constructed from our results. 

%%%%%%%%%%%%%%%%%%%%%%%%%%%
\section{Semiclassical scattering amplitudes}\label{sec:intro-to-semi}

In standard perturbation theory around a trivial vacuum, tree-level scattering amplitudes can be given a purely variational definition, as multi-linear pieces of the classical action evaluated on recursively constructed solutions to the equations of motion. The order to which one constructs the solution perturbatively in the coupling and its boundary conditions are dictated by the multiplicity of the scattering process and the asymptotic quantum numbers of the scattered states, respectively. This is sometimes called the \emph{perturbiner approach} to scattering amplitudes~\cite{DeWitt:1967ub,Arefeva:1974jv,Abbott:1983zw,Jevicki:1987ax,Rosly:1996cp,Rosly:1996vr,Rosly:1997ap,Selivanov:1999as,Mizera:2018jbh,Garozzo:2018uzj,Cho:2021nim,Lee:2022aiu,Cho:2022faq}, which trades the combinatorial computations of traditional Feynman rules for computations in differential equations and variational calculus. 

The perturbiner approach can easily be extended to scattering amplitudes in background (gauge and gravitational) fields by extracting multilinear pieces of the classical background field action (cf., \cite{Gorsky:2005sf,Mason:2009afn,Adamo:2017nia,Adamo:2020qru,Adamo:2021rfq})\footnote{An equivalent definition of these quantities as `on-shell' correlators has been given in~\cite{Meltzer:2020qbr,Sivaramakrishnan:2021srm,Cheung:2022pdk}}. When the background fields admit an S-matrix (e.g., ultrarelativistic beams, shockwaves and sandwich plane waves in gauge theory and gravity), the amplitudes obtained from the perturbiner approach agree with those obtained using background-coupled Feynman rules. However, even when the S-matrix does \emph{not} exist -- as in black hole spacetimes~\cite{DeWitt:1953ipa,Hawking:1974rv,Hawking:1975vcx,Gibbons:1975kk,Gibbons:1975jb,Woodhouse:1976fe} -- the perturbiner approach remains well-defined. The resulting amplitudes are gauge-invariant quantities which contain all of the dynamical information expected from tree-level background field scattering amplitudes which is needed to compute observable quantities; see~\cite{Ilderton:2023ifn} for explicit examples in the case of electromagnetic horizons. As such we continue to use to word `amplitudes' for the output of perturbiner calculations. Other potential ambiguities associated with a particular backgrounds (e.g., lack of a unique choice of vacuum) will manifest themselves in the choices of admissible boundary conditions for the background-coupled equations of motion.

The recursive construction of solutions to the equations of motion in the perturbiner approach is seeded with solutions to the free equations of motion with boundary conditions corresponding to asymptotic scattering states; this is simply the perturbiner version of LSZ truncation. The external states corresponding to the full tree-level S-matrix (which will include quantum information when there are massive particles involved) are thus exact solutions to the `free' equations of motion on the background. However, on many backgrounds including Coulomb potentials in QED and black holes in general relativity (linearised or fully non-linear), the required solutions are so complicated that explicit calculations of scattering amplitudes -- particularly in the presence of emitted radiation -- have simply not been possible.

To address this, we introduce here a tractable approximation of tree-level scattering amplitudes in background fields which we refer to as \emph{semiclassical scattering amplitudes}. These semiclassical amplitudes are defined by taking the semiclassical WKB approximation for the external states in the scattering process -- that is, by approximating solutions to the free equations of motion in the background -- and using these as the input for the perturbiner approach. To be precise: 

\begin{definition}[Semiclassical scattering amplitude]\label{def:sca}
A tree-level scattering amplitude (in the sense of the perturbiner approach) with all external legs defined by solving the free equations of motion using the WKB expansion to leading order in the $\hbar\to0$ limit.
\end{definition}

For external massless fields, where $\hbar$ does not enter the equations of motion, this semiclassical prescription is not an approximation: the free equation of motion is classically exact. Thus, massless legs in a semiclassical scattering amplitude are represented by solutions to their full equation of motion, without approximation.

As we will see, semiclassical amplitudes defined in this way are controlled by Hamilton's principal function (HPF): the solution to the Hamilton-Jacobi equations for a particle on the background. Even when there are massless external legs which interact with the background -- as in the emission of gravitational radiation on a curved spacetime -- this remains the case, as the graviton wavefunction can be written in terms of the HPF and a background-dressed polarization tensor, which is itself related to the HPF through the linearised Einstein equation.

The massive free field equations on the asymptotically flat, static and spherically symmetric backgrounds (i.e., Coulomb and Schwarzschild) that we consider in this paper are \emph{not} WKB exact, so these semiclassical amplitudes contain less information than the full tree-level S-matrix, but -- as we will see -- they encode the underlying classical probe dynamics in the background and its weak field limit.

%%%%%%%%%%%%%%%%%%%%%%%%%%%
%%%%%%%%%%%%%%%%%%%%%%%%%%%

\section{Semiclassical scalar wavefunctions \& elastic scattering}\label{sec:2pt}

The external wavefunctions for any scattering process in a background are defined by solving the free, background-coupled equations of motion for the asymptotic incoming and outgoing states involved in the process. For massive scalar particles coupled to electromagnetism or gravity, these are solutions to the Klein-Gordon equation on the given background with appropriate boundary conditions. The standard approach to solving these equations for spherically symmetric backgrounds like a Coulomb field or the Schwarzschild metric is to separate variables, reducing the problem to a second-order radial ODE for the coefficient functions of a spherical harmonic expansion (cf., \cite{Darwin:1913,Regge:1957td,Dollard:1964,Zerilli:1970se,Taylor:1973da,Chandrasekhar:1985kt,Glampedakis:2001cx,Boyer:2004,Kol:2021jjc}). The 2-point (or, more properly, $1\to1$) amplitude for elastic scattering is then read off -- at least, in principle -- from the asymptotic expansion of these radial wavefunctions (cf., \cite{Taylor:1972pty,Itzykson:1980rh,Futterman:1988ni,Landau:1991wop}). Simplifications arise under assumptions such as small momentum exchange,
ultrarelativistic limits or a perturbative description of the background (e.g., \cite{tHooft:1987vrq,Jackiw:1991ck,Kabat:1992tb,Lodone:2009qe,Adamo:2021jxz,Adamo:2021rfq}).

Here, we show that semiclassical scalar wavefunctions in static, spherically symmetric backgrounds can be determined (at all-orders in the coupling, for generic scattering angle) by using a WKB ansatz combined with an asymptotic matching condition to the usual radial wavefunctions. This procedure is inspired by a similar `patching' approach to solving the Klein-Gordon equation in the WKB-exact, ultrarelativistic setting where the background is localized on a lightfront~\cite{Penrose:1965rx,tHooft:1987vrq,Klimcik:1988az,Adamo:2021jxz}, suitably generalized to non-WKB-exact backgrounds like Coulomb and Schwarzschild. States constructed in this way have the substantial advantage of being highly amenable to calculation in the relativistic, covariant framework of background QFT. As a warm-up, we show how the 2-point, elastic scattering amplitudes for semiclassical scalars are obtained in a straightforward way using these states.  

%%%%%%%%%%%%%%%%%%%%%%%%%%%%%%%%%%%%%%%%%%%%%

\subsection{Semiclassical scalar states on static, spherically symmetric backgrounds}

On-shell complex scalar fields coupled to background electromagnetic or gravitational fields are defined by solutions to the Klein-Gordon equation in the background. Let $A_{\mu}(x)$ denote a background electromagnetic field solving Maxwell's equations in Minkowski spacetime and $g_{\mu\nu}(x)$ denote a background metric solving the Einstein equations. Working in Lorenz gauge for the electromagnetic background ($\partial^{\mu}A_{\mu}=0$) and de Donder gauge for the gravitational background ($g^{\mu\nu}\,\Gamma^{\alpha}_{\mu\nu}=0$ for $\Gamma^{\alpha}_{\mu\nu}$ the Christoffel symbols of $g_{\mu\nu}$), the Klein-Gordon equations become
\be\label{KG1}
\left(\Box-\frac{2\im}{\hbar}\,e\,A^{\mu}\partial_{\mu}-\frac{e^2}{\hbar^2}\,A^2+\frac{m^2}{\hbar^2}\right)\phi(x)=0\,, \qquad \left(g^{\mu\nu}\partial_{\mu}\partial_{\nu}-\frac{m^2}{\hbar^2}\right)\phi(x)=0\,.
\ee
In the first equation, $\Box:=\eta^{\mu\nu}\partial_{\mu}\partial_{\nu}$ is the Minkowski spacetime d'Alembertian, $e$ is the charge of the complex scalar and all indices are contracted via the Minkowski metric. 

To define semiclassical states from solutions to these equations, we employ a WKB approximation
\be\label{WKB1}
\phi(x)=\e^{\im\,\frac{S(x)}{\hbar}}\,.
\ee
In the $\hbar\to0$ semiclassical limit, the Klein-Gordon equations \eqref{KG1} become the Hamilton-Jacobi equations for the background:
\be\label{HJ1}
\eta^{\mu\nu}\,\left(\partial_\mu S- e\,A_{\mu}\right)\,\left(\partial_\nu S- e\,A_{\nu}\right)=m^2\,,
\qquad g^{\mu\nu}\,\partial_{\mu}S\,\partial_{\nu}S=m^2\,.
\ee
In other words, in the semiclassical limit the field equations impose that the WKB phase $S(x)$ becomes Hamilton's principal function (HPF) for the background. From now on, we will set $\hbar=1$ in most expressions, with the implicit understanding that we work in the semiclassical limit described by \eqref{HJ1}; the dropping of quantum contributions will be highlighted where necessary.

Now, to obtain scattering states parametrized by an on-shell asymptotic momentum $p_{\mu}$ obeying $\eta^{\mu\nu}p_{\mu}p_{\nu}=m^2$, we follow~\cite{Adamo:2021rfq} and make a weak-field expansion of the WKB phase
\be\label{Phaseexp}
S(x)=\sum_{n=0}^{\infty}S^{(n)}(x)\,, \qquad S^{(n)}(x)\propto e^n,\,G^n\,,
\ee
where $e$ is the elementary charge in the electromagnetic case and $G$ is Newton's constant in the gravitationally-coupled case. We also assume the existence of a similar weak-field expansion of the background fields themselves; for the inherently linear electromagnetic background this is trivial, while for gravity it implies
\be\label{BFexp}
g_{\mu\nu}=\eta_{\mu\nu}+H_{\mu\nu}+\sum_{n=2}^{\infty}H^{(n)}_{\mu\nu}\,
\ee
with $H_{\mu\nu}\propto G$ the leading, linear correction to Minkowski spacetime and $H^{(n)}_{\mu\nu}\propto G^n$ encoding the higher, non-linear terms.

This leads to a system of coupled differential equations at each order in the expansion:
\be\label{phases0}
\partial_{\mu}S^{(0)}\,\partial^{\mu}S^{(0)}=m^2\,,
\ee
\be\label{phases1}
\partial^{\mu}S^{(0)}\,\partial_{\mu}S^{(1)}= e\,A^{\mu}\,\partial_{\mu}S^{(0)}\,,
\qquad
2\,\partial^{\mu}S^{(0)}\,\partial_{\mu}S^{(1)}=H^{\mu\nu}\,\partial_{\mu}S^{(0)}\,\partial_{\nu}S^{(0)}\,,
\ee
to subleading order in the weak field expansion, with all indices in all equations now contracted using the Minkowski metric. The (theory-independent) leading equation for $S^{(0)}$ can then be solved in terms of an on-shell momentum $p_{\mu}$:
\be\label{S0}
S^{(0)}(x)=p\cdot x\,, \qquad p^2=\eta^{\mu\nu}p_{\mu}\,p_{\nu}=m^2\,,
\ee
in Minkowski spacetime.

We now make an additional simplifying assumption, which in effect restricts us to the cases of interest: we assume that the background field is spherically symmetric. By Birkhoff's theorems in both electromagnetism~\cite{Pappas:1984} and general relativity~\cite{Jebsen:1921,Birkhoff:1923,Israel:1967wq} this implies that the backgrounds are static and asymptotically flat. In particular, if we assume that scattering occurs in a vacuum region of spacetime (i.e., outside of any sources), the electromagnetic and gravitational background fields are the Coulomb gauge potential and Schwarzschild metric, respectively. In terms of the fields entering the PDEs defining $S^{(1)}$ in \eqref{phases1}, we have, in spherical polar coordinates $(t,r,\theta,\phi)$
\be\label{CoulSchw}
A_{\mu}=\frac{Q\,U_{\mu}}{4\,\pi\,r}\,, \qquad H_{\mu\nu}=\frac{2\,G\,\cP_{\mu\nu}}{r}\,,
\ee
where $Q$ is the charge of the Coulomb potential, $U_{\mu}=(1,0,0,0)$ and $\cP_{\mu\nu}$ is the constant tensor
\be\label{projector}
\cP_{\mu\nu}:=M\left(\eta_{\mu\alpha}\,\eta_{\nu\beta}+\eta_{\mu\beta}\,\eta_{\nu\alpha}-\eta_{\mu\nu}\,\eta_{\alpha\beta}\right)U^{\alpha}\,U^{\beta}\,,
\ee
for $M$ the Schwarzschild mass. The equations \eqref{phases1} for $S^{(1)}$ are then easily integrated to give
\be\label{S1em}
S^{(1)}(x)=\frac{\,eQ\,p\cdot U}{4\,\pi\,|\vec{p}|}\,\log(|\vec{p}|r+\vec{p}\cdot\vec{r})\,,
\ee
for electromagnetism and 
\be\label{S1gr}
S^{(1)}(x)=\frac{G\,\cP^{\mu\nu}\,p_{\mu}\,p_{\nu}}{|\vec{p}|}\,\log(|\vec{p}|r+\vec{p}\cdot\vec{r})\,
\ee
for gravity. In both expressions, $\vec{p}$ denotes the spatial components on the on-shell momentum $p_{\mu}$ with (Euclidean) norm $|\vec{p}|$.

It should be noted that in the context of large-distance, the phase $S^{(1)}$ is equal to an \emph{eikonal phase}, which is a function of the radial distance $r$ alone (cf., \cite{Kabat:1992tb,Kol:2021jjc,Adamo:2021rfq}). However, if one wishes to consider situations beyond elastic 2-point scattering, the non-trivial angular dependence of $S^{(1)}$ is crucial, as we will show later.

\medskip

More generally, one can proceed recursively to solve for the HPF \eqref{Phaseexp} order-by-order in the weak-field expansion. Let $S_{p}(x)$ denote the resulting \emph{all-order} HPF corresponding to on-shell asymptotic momentum $p_{\mu}$; a general solution to the free field equations \eqref{KG1} can then be determined by taking an on-shell linear combination of such particular solutions:
\be\label{gensoln}
\phi(x)=\int \d\Phi(p)\,\Lambda(p)\,\e^{\im\,S_p(x)}\,,
\ee
where (defining ${\hat \ud}^4 p \equiv \ud^4p/(2\pi)^4$ and $\hat\delta(\cdot) = 2\pi \delta(\cdot)$), 
\be\label{LIPS}
\d\Phi(p):={\hat\d}^{4}p\,\,\Theta(p^0)\,\hat\delta(p^2-m^2)\,,
\ee
is the Lorentz-invariant on-shell measure and $\Lambda(p)$ are as-yet-undetermined coefficients.

At this point, we take inspiration from a similar procedure for constructing general solutions to the Klein-Gordon equation on plane- or pp-wave backgrounds which are localized on a lightfront~\cite{Penrose:1965rx,tHooft:1987vrq,Klimcik:1988az,Adamo:2021jxz}. In that context, the coefficients in the on-shell superposition (\ref{gensoln}) are determined by matching conditions at this lightfront, namely that the equation of motion is solved on the lightfront itself. Such backgrounds which include impulsive plane waves, ultrarelativistic shockwaves and beams are, of course, very different from Coulomb or Schwarzschild: they are WKB exact, so the procedure determines fully quantum mechanical scattering states.
The lesson we wish to apply to the context of \eqref{gensoln} in a Coulomb or Schwarzschild background is that the coefficients in the sum can be fixed by demanding that the solution has desired properties in a certain region of spacetime. 

In Coulomb or Schwarzschild, the natural matching region is at spatial infinity\footnote{In spherical polar coordinates, $r\to\infty$ is the asymptotic boundary region associated to the Coulomb potential and Schwarzschild metric. However, one could instead write the background fields in some other coordinate system with a different boundary; for instance, retarded Bondi coordinates would give future null infinity $\scri^+$ as the natural boundary. The matching conditions and wavefunctions in such alternative coordinates will certainly look very different to those in spherical coordinates, but diffeomorphism invariance ensures that the corresponding scatting amplitudes themselves will agree; see~\cite{Fabbrichesi:1993kz,Gonzo:2022tjm} for some 2-point examples obtained from $\scri$.}, $r\to\infty$. Here, our solution should agree with solutions to the Klein-Gordon equation obtained in the `standard' way, by separating variables and expanding in spherical harmonic modes. The radial wavefunctions obtained in this way encode the elastic scattering amplitude in their asymptotic behaviour as $r\to\infty$, so clearly the solutions \eqref{gensoln} must agree with them in this asymptotic region. This is enough to fix the coefficients $\Lambda(p)$.

\medskip

To begin, recall that general solutions to the Klein-Gordon equations \eqref{KG1} in a static, spherically symmetric background can be written in terms of a spherical harmonic expansion
\be\label{shexp}
\phi_p(x)=\frac{4\pi\,\e^{\im\,E\,t}}{r}\,\sum_{\ell=0}^{\infty}\sum_{m=-\ell}^{\ell}Y_{\ell}^{m}(\hat{x})\,\overline{Y_{\ell}^{m}}(\hat{p})\,R_{\ell m}(r)\,,
\ee
where $E\equiv p^0$, $Y_{\ell}^{m}(\hat{x})$ are the usual spherical harmonics evaluated at $(\theta,\varphi)$ on the unit sphere $\hat{x}=(\sin\theta\,\cos\varphi,\,\sin\theta\,\sin\varphi,\,\cos\varphi)$ and $\hat{p}=\vec{p}/|\vec{p}|$ is the unit vector associated to the spatial momentum. The non-trivial part comes from solving for the radial wavefunction modes $R_{\ell m}(r)$, which are determined by a second-order ODE of Schr\"odinger type. The functional form of these radial wavefunctions can be quite complicated: in Coulomb or linearized Schwarzschild backgrounds they are Whittaker/confluent hypergeometric functions, while in a fully non-linear Schwarzschild metric they are confluent Heun functions. Luckily, we will only require the asymptotic form of the radial wavefunctions as $r\to\infty$, and their asymptotic expansions are well-known (cf., \cite{NIST:DLMF} Sections 33 and 31, respectively).  

First, we consider the asymptotic expansion of the semiclassical WKB ansatz. For $S^{(0)}=p\cdot x$ one invokes the plane wave expansion and spherical harmonic addition theorem to find
\be\label{pwexp}
\e^{\im\,p\cdot x}=4\pi\,\sum_{\ell=0}^{\infty}\sum_{m=-\ell}^{\ell}\im^{\ell}\,j_{\ell}(|\vec{p}|r)\,Y_{\ell}^{m}(-\hat{x})\,\overline{Y_{\ell}^{m}}(\hat{p})\,\e^{\im\,E\,t}\,,
\ee
where $j_\ell$ are the spherical Bessel functions. As $r\to\infty$ these obey
\be\label{bessexp}
j_{\ell}(|\vec{p}|r)=\frac{\e^{\im\,(|\vec{p}|r-\ell\,\pi/2)}}{2\,\im\,|\vec{p}|\,r}-\frac{\e^{-\im\,(|\vec{p}|r-\ell\,\pi/2)}}{2\,\im\,|\vec{p}|\,r}+O(r^{-2})\,.
\ee
The two leading terms correspond to left or right-moving particles in the semiclassical state, respectively; only the second term is relevant for the desired state in the on-shell sum \eqref{gensoln}, so the first term is discarded by hand\footnote{If we proceeded na\"ively, keeping the first term leads to an apparent un-physical divergence in the wavefunction. To see that the first term corresponds to a \emph{finite} contribution which simply has the wrong scattering behaviour requires a more careful treatment of all harmonic and asymptotic expansions, which is described in Appendix~\ref{app:Asymp}.}.

From \eqref{S1em} -- \eqref{S1gr} it is clear that the next correction to the HPF, $S^{(1)}(x)$ grows like $\log r$ as $r\to\infty$, while all terms in the HPF $S^{(n)}$ for $n\geq2$ scale as $O(r^{-1})$ (see~\cite{Akhoury:2013yua,Bjerrum-Bohr:2016hpa,Luna:2016idw,KoemansCollado:2019ggb} for explicit expressions at $n=2$ in the eikonal regime); this follows simply by inspection of the structure of the Hamilton-Jacobi equations at higher-orders in the weak-field expansion. Thus, it follows that the semiclassical WKB wavefunction, to all orders in the coupling, behaves at large $r$ as
\be\label{WKBexp1}
\e^{\im\,S_p(x)}=\frac{2\pi\,\im}{|\vec{p}|\,r}\left(\sum_{\ell=0}^{\infty}Y_{\ell}^{m}(\hat{x})\,\overline{Y_{\ell}^{m}}(\hat{p})\right)\,\e^{\im\left(E\,t-|\vec{p}|\,r+S^{(1)}(x)\right)}+O(r^{-2})\,,
\ee
for an outgoing state. Using the spherical harmonic completeness relations, this is further simplified to
\be\label{WKBexp2}
\e^{\im\,S_p(x)}=\frac{2\pi\,\im}{|\vec{p}|\,r}\,\delta^{2}_{\Omega_x}(\hat{x}-\hat{p})\,\e^{\im\left(E\,t-|\vec{p}|\,r+S^{(1)}(x)\right)}+O(r^{-2})\,,
\ee
where
\be\label{spheredelt}
\delta^{2}_{\Omega_x}(\hat{x}-\hat{p}):=\frac{1}{\sin\theta}\,\delta^{2}(\hat{x}-\hat{p})\,,
\ee
covariantly localizes the angular dependence to that of the on-shell momentum. On the support of these delta functions
\be\label{S1asymp}
S^{(1)}(x)\longrightarrow C_p\,\log(2|\vec{p}|\,r)\,,
\ee
for $C_p$ the theory-dependent constant pre-factor determined by \eqref{S1em} and \eqref{S1gr} in electromagnetism and gravity, respectively.

Feeding \eqref{WKBexp2} and \eqref{S1asymp} into \eqref{gensoln}, the asymptotic behaviour of the general solution is
\be\label{gensolasymp}
\phi(x)=\frac{2\pi\im}{E\,r}\,\int_{0}^{\infty}|\vec{p}|\,\d|\vec{p}|\,\Lambda(|\vec{p}|,\hat{x})\,\e^{\im\left(E\,t-|\vec{p}|\,r+C_p\log(2|\vec{p}|\,r)\right)}+O(r^{-2})\,,
\ee
where three of the four on-shell phase space integrals have been done trivially against delta functions. 

At this point, the asymptotic matching condition between \eqref{gensolasymp} and an exact solution of the form \eqref{shexp} with momentum $p'_{\mu}$ and energy $E$ reads:
\begin{equation}\label{matching1}
\begin{split}
\frac{\im}{E}\int_{0}^{\infty}|\vec{p}|\,\d|\vec{p}|\,\Lambda^{p'}(|\vec{p}|,\hat{x})\, &\lim_{r\to\infty}\e^{\im\left(E\,t-|\vec{p}|\,r+C_p\log(2|\vec{p}|\,r)\right)} \\
&=2\,\e^{\im\,E\,t}\,\sum_{\ell=0}^{\infty}\sum_{m=-\ell}^{\ell}Y_{\ell}^{m}(\hat{x})\,\overline{Y_{\ell}^{m}}(\hat{p}\,^{\prime})\,\lim_{r\to\infty}R_{\ell m}(r)\,,
\end{split}
\end{equation}
where the superscript on $\Lambda^{p'}$ denotes the fact that these coefficients are being fixed by matching with a solution of momentum $p'$. This condition can be solved by taking
\begin{multline}\label{Lambda1}
\Lambda^{p'}(|\vec{p}|,\hat{p})=-\frac{2\im\,E}{|\vec{p}|}\,\delta(|\vec{p}|-|\vec{p}\,^{\prime}|) \\
\times\,\sum_{\ell=0}^{\infty}\sum_{m=-\ell}^{\ell}Y_{\ell}^{m}(\hat{p})\,\overline{Y_{\ell}^{m}}(\hat{p}\,^{\prime})\,\lim_{r\to\infty}R_{\ell m}(r)\,\e^{\im\left(|\vec{p}|\,r-C_p\log(2|\vec{p}|\,r)\right)}\,.
\end{multline}
Now, as $r\to\infty$ the radial wavefunctions behave at leading order as (cf., \cite{Kol:2021jjc}):
\be\label{RadialWF}
R_{\ell m}(r)\xrightarrow{r\to\infty} \e^{-\im\left(|\vec{p}|\,r-C_p\log(2|\vec{p}|\,r)\right)+\im\,B_{\ell}}\,,
\ee
where $B_{\ell}$ depend on the kinematics and mode number $\ell$ but are otherwise constant. In other words, the $r$-dependence appearing in the large-$r$ limit part of the matching condition \eqref{Lambda1} precisely cancels, leaving a finite result.

The combination of spherical harmonics and large-$r$ limits appearing in \eqref{Lambda1} can then be repackaged into a (finite) partial wave sum known as the elastic scattering amplitude~\cite{Kol:2021jjc}: 
\be\label{elamp}
\begin{split}
f^{p'}(|\vec{p}|,\hat{p})&:=\sum_{\ell=0}^{\infty}\sum_{m=-\ell}^{\ell}Y_{\ell}^{m}(\hat{p})\,\overline{Y_{\ell}^{m}}(\hat{p}\,^{\prime})\,\lim_{r\to\infty}R_{\ell m}(r)\,\e^{\im\left(|\vec{p}|\,r-C_p\log(2|\vec{p}|\,r)\right)} \\
 &=\sum_{\ell=0}^{\infty}(2\ell+1)\,P_{\ell}(\hat{p}\cdot\hat{p}\,^{\prime})\,\e^{2\im\,I_{\ell}(r=\infty)-\im\pi\ell}\,,
\end{split}
\ee
where $P_{\ell}$ are the Legendre polynomials and $I_{\ell}(r=\infty)$ is the \emph{radial action}~\cite{Landau:1976} at infinity, regularized by the subtraction of a divergent accumulation phase. The radial action takes the explicit forms (cf., \cite{Kol:2021jjc})
\be\label{CoulRA}
I^{\mathrm{Coulomb}}_\ell(r)=\int_{r_{\text{turn}}}^{r} \sqrt{\frac{\vec{p}\,^2\, s^2+\frac{E\,eQ}{2\pi}\,s-\nu_{\ell}^{2}}{s^2}}\, \d s \,,
\ee
for $\nu_\ell=\sqrt{\ell^2-\frac{eQ}{4\pi}}-1/2$ in Coulomb and
\be\label{SchRA}
I^{\mathrm{Sch}}_\ell(r)=\int_{r_{\text{turn}}}^{r} \sqrt{\frac{\vec{p}\,^2\,s^2+2GMm\, s-\frac{s-2 G M}{s} \ell^2}{(s-2 G M)^2}}\, \d s \ .
\ee
in Schwarzschild, and the accumulation phase takes the universal form $|\vec{p}|r+\eta\,\log(2|\vec{p}|r)$ for
\be\label{apparam}
\eta^{\mathrm{Coulomb}}:=\frac{E\,e\,Q}{2\pi\,|\vec{p}|}\,, \qquad \eta^{\mathrm{Sch}}:=\frac{G\,M}{|\vec{p}|}\left(E^2+\vec{p}\,^{2}\right)\,.
\ee
In both cases, $r_{\text{turn}}$ is the value of $s$ for which the integrand -- the radial momentum of a freely falling probe -- vanishes.

It is worth mentioning that in the limit where the angle between $\hat{p}$ and $\hat{p}\,^{\prime}$ is small, it is well-known that the partial wave sum in \eqref{elamp} can be expressed as an eikonal integral (cf., \cite{Bethe:1958zz,Berry:1972na,Jauch:1976ava,Ford:2000uye,Glampedakis:2001cx,Gribov:2003nw,Bautista:2021wfy}). In this eikonal limit, the coefficients $\Lambda(p)$ are still given by \eqref{Lambda2}, but the amplitude $f$ is now given by
\begin{equation}\label{eq:lambda-match}
    f^{p'}(|\vec{p}|,\,\hat{p})=\im\,|\vec{p}|\: \int \d^2x^{\perp}\, \e^{\im\,x^{\perp}\cdot(\hat{p}-\hat{p}\,^{\prime})} \,\e^{\im\left(2I(|x^{\perp}|)-\pi\,|\vec{p}|\,|x^{\perp}|\right)}\,,
\end{equation}
where 
\be\label{eikphase}
I(|x^{\perp}|):=I_{|\vec{p}|\,|x^{\perp}|-\frac{1}{2}}(r=\infty)\,,
\ee
in terms of the radial action. This is due to the eikonal limit of the partial wave sum being dominated by large $\ell$ contributions, and the eikonal integral \eqref{eq:lambda-match} is similarly dominated by a large $|x^{\perp}|$ saddle point~\cite{Abarbanel:1969ek,Levy:1969cr,Amati:1987uf,Ciafaloni:2015xsr}.

\medskip

To summarize, for arbitrary scattering angle the matching condition fixes
\be\label{Lambda2}
\Lambda^{p'}(|\vec{p}|,\hat{p})=-\frac{2\im\,E}{|\vec{p}|}\,\delta(|\vec{p}|-|\vec{p}\,^{\prime}|)\,f^{p'}\!(|\vec{p}|,\,\hat{p})\,,
\ee
with $f^{p'}(|\vec{p}|,\,\hat{p})$ determined by the radial action of the background. Feeding this back into the initial form of the general solution \eqref{gensoln} gives the final expression for an outgoing state associated with momentum $p^{\mu}$:
\be\label{gensolfin}
\begin{split}
\phi_{p}(x)&=\int \d\Phi(l)\,\Lambda^{p}(l)\,\e^{\im\,S_{l}(x)}\Big|_{l^0=p^0} \\
 & =-\im\,|\vec{p}|\,\int \d^{2}\Omega_{l}\,f^{p}(|\vec{p}|,\,\hat{l})\,\e^{\im\,S_{l}(x)}\Big|_{l^0=p^0}\,,
\end{split}
\ee
upon performing two of the on-shell phase space integrals. The remaining integral is over the celestial sphere with measure 
\be\label{spheremeasure}
\d^{2}\Omega_{l}:=\sin\theta_{\hat{\ell}}\,\d\theta_{\hat{l}}\,\d\varphi_{\hat{l}}\,, 
\ee 
corresponding to the angles defined by $\hat{l}$. 

%%%%%%%%%%%%%%%%%%%%%%%%%%

\subsection{Scattering without emission from the radial action}
\label{sec:all-from-radial}
Usually, the elastic $1\to1$ scattering amplitude in a static, spherically symmetric background is read off from the asymptotic expansion of the radial wavefunction, with the results for Coulomb and Schwarzschild backgrounds being well-known~\cite{Darwin:1913,Regge:1957td,Zerilli:1970se,Futterman:1988ni,Glampedakis:2001cx,Boyer:2004,Kol:2021jjc}. Here, we show how these results can be obtained directly from the perturbiner approach described in Sec.~\ref{sec:intro-to-semi}.
This is important for two reasons: firstly, it extends the results of~\cite{Adamo:2021rfq} connecting $1\to1$ scattering in linearised Schwarzschild with the eikonal amplitude to $1\to1$ scattering in the \emph{exact, non-linear} Schwarzschild metric, and secondly, it serves as a useful warm-up for the case of $1\to2$ scattering with emission that we will consider in subsequent sections.

The 2-point amplitude in this framework is given by the quadratic part of the background coupled classical action, evaluated on the sum of an incoming and outgoing state, taking only the contribution linear in each state. The fact that these states solve the free equation of motion means that only a boundary term contributes\footnote{For the Schwarzschild metric, there is another boundary at $r=2GM$, the event horizon. By ignoring this boundary's contributions to the 2-point amplitude, we are implicitly considering elastic scattering with sufficiently large impact parameter that the probe's interaction with the horizon is vanishingly small. This assumption is implicit in most considerations of $1\to1$ scattering in black hole spacetimes, although there are several interesting studies that consider horizon effects in different frameworks (e.g., \cite{tHooft:1996rdg,Betzios:2016yaq,Goldberger:2019sya,Goldberger:2020geb,Goldberger:2020wbx,Betzios:2020xuj,Kallosh:2021ors,Gaddam:2021zka}).}, and for background Coulomb of Schwarzschild fields (expressed in spherical coordinates) it is straightforward to show that this is given by 
\begin{equation}\label{eq:2-point-r}
    \bra{p'}\mathcal{S}\ket{p}:= \lim_{r\rightarrow\infty}
    \int_{\R\times S^2} \d t\,\d^{2}\Omega_x \, r^2\, \bar{\phi}^{\mathrm{in}}_{p}\,\partial_{r}\phi^{\mathrm{out}}_{p'}\,,
\end{equation}
for $1\to1$ scattering of a charged or gravitationally-coupled complex scalar with initial momentum $p$ and final momentum $p'$. As the amplitude is totally localised at spatial infinity, the scattering conditions at this boundary are that the incoming and outgoing states appearing in \eqref{eq:2-point-r} are given by
\be\label{2pstates}
\phi^{\mathrm{in}}_{p}(x)=\e^{\im\,S_{p}(x)}\,, \qquad \phi^{\mathrm{out}}_{p'}(x)=-\im\,|\vec{p}\,^{\prime}|\,\int \d^{2}\Omega_{l}\,f^{p'}\!(|\vec{p}\,^{\prime}|,\,\hat{l})\,\e^{\im\,S_{l}(x)}\Big|_{l^0=(p')^0}\,,
\ee
in terms of the HPF and background dressing \eqref{gensolfin}.

Plugging these into \eqref{eq:2-point-r} and exploiting the asymptotic expansion \eqref{WKBexp2} gives
\begin{multline}\label{2ptc1}
\bra{p'}\mathcal{S}\ket{p}=\frac{|\vec{p}\,^{\prime}|}{|\vec{p}|}\,\lim_{r\to\infty}\int\limits_{\R\times(S^2)^2}\d t\,\d^{2}\Omega_x\,\d^{2}\Omega_l\,\delta^{2}_{\Omega_x}(\hat{x}-\hat{p})\,\delta^{2}_{\Omega_x}(\hat{x}-\hat{l})\,f^{p'}\!(|\vec{p}\,^{\prime}|,\,\hat{l}) \\
\times \exp\!\left[\im\left((E'-E)\,t+(|\vec{p}|-|\vec{p}\,^{\prime}|)\,r+C_{p'}\,\log(2|\vec{p}\,^{\prime}|r)-C_{p}\,\log(2|\vec{p}|r)\right)\right]\,,
\end{multline}
dropping all terms which vanish in the $r\to\infty$ limit. The time integral is now performed to give a delta function setting $E'=E$, and thus $|\vec{p}|=|\vec{p}\,^{\prime}|$, as the masses of the incoming and outgoing states are equal. This immediately removes all remaining $r$-dependence from the integrand, rendering the $r\to\infty$ limit trivial and leaving 
\be\label{2ptc2}
\begin{split}
\bra{p'}\mathcal{S}\ket{p}&=\hat{\delta}(E'-E)\,\int\limits_{S^2\times S^2}\d^{2}\Omega_x\,\d^{2}\Omega_l\,\delta^{2}_{\Omega_x}(\hat{x}-\hat{p})\,\delta^{2}_{\Omega_x}(\hat{x}-\hat{l})\,f^{p'}\!(|\vec{p}|,\,\hat{l}) \\
 &=\hat{\delta}(E'-E)\,f^{p'}\!(|\vec{p}|,\,\hat{p})\,.
\end{split}
\ee
In particular, we find that the well-known result that the $1\to1$ elastic scattering amplitude is given by $f^{p'}\!(|\vec{p}|,\,\hat{p})$. By \eqref{elamp}, this also confirms that the $1\to1$ amplitude on any spherically symmetric background is captured exactly, without any small angle or leading-order eikonal approximation, by the full outgoing amplitude, which is itself controlled by the radial action associated with the background~\cite{Kol:2021jjc}.

\medskip

\paragraph{The scattering angle:} It is interesting to see how classical physics, such as the scattering angle, is encoded in the 2-point amplitude \eqref{2ptc2}. The Legendre polynomials appearing in \eqref{elamp} describe the angular dependence of the wavefunction for the outgoing state and contain both classical and quantum information; in keeping with the correspondence principle~\cite{Bohr:1976}, the classical information is encoded only by certain values of $\ell$ in the partial wave sum, namely (cf., \cite{Landau:1991wop}, Chapter VII \S49 and XVII \S127):
\begin{equation}
    \theta\, \ell \gg 1\,, \qquad  \left(\pi-\theta\right)\ell \gg1 \,,
\end{equation}
for $\cos\theta=\hat{p}\cdot\hat{p}\,^{\prime}$ (not to be confused with the spherical polar coordinate $\theta$). The condition of classicality for the angular part of the wavefunction can thus be expressed as: for a given value of $\theta\neq0,\pi$, the classical limit occurs for large values of $\ell$, which corresponds to a small variation in the De Broglie wavelength. In this limit, the Legendre polynomials can be expanded as
\begin{equation}\label{eq:app-Leg}
    P_{\ell}(\cos\theta)=\sqrt{\frac{\theta}{\sin\theta}}\,J_{0}\!\left(\left(\ell+\frac{1}{2}\right)\theta\right)+O((\theta \ell)^{-3/2})\,, \qquad \ell \gg \theta^{-1}\,,
\end{equation}
where $J_0$ is the Bessel function of the first kind. Substituting this expansion into \eqref{elamp} does not result in any loss of information so long as the resulting expression is understood in the classical (i.e., large $\ell$) limit. 

In particular, this eliminates the need to use a small-angle approximation to arrive at the expansion \eqref{eq:app-Leg}, as was done in~\cite{Bautista:2021wfy}. One then defines
\be\label{bdef}
b:=\frac{(\ell+1/2)}{|\vec{p}\,^{\prime}|}\,,
\ee
so that the high-energy limit is now equivalent to assuming small variations of $b$ and the sum over $\ell$ can be replaced with an integral over $b$. The $2$-point amplitude then becomes
\begin{equation}
\begin{aligned}\label{eq:2spl}
    \bra{p'}\mathcal{S}\ket{p}&= 2\im\,\hat{\delta}(E'-E)\,|\vec{p}|\,\sqrt{\frac{\theta}{\sin\theta}}\,\int_{0}^{\infty} b\,\d b\,J_0(|\vec{p}|\,\theta\,b)\,\e^{\im\left(2\,I(b)-\pi\,|\vec{p}|\,b\right)} \\
&=\frac{\im}{\pi}\,\hat{\delta}(E'-E)\,|\vec{p}|\,\sqrt{\frac{\theta}{\sin\theta}}\,\int_{0}^{\infty} b\,\d b\,\int_{0}^{2\pi}\d\varphi\,\e^{\im\,|\vec{p}|\,\theta\,b\,\cos\varphi}\,\e^{\im\left(2\,I(b)-\pi\,|\vec{p}|\,b\right)}\,,
     \end{aligned}
\end{equation}
where $I(b)$ is defined in \eqref{eikphase} and the last line follows from the integral representation of the Bessel function.  Performing all integrals via saddle point approximation, one finds that the integral is sharply peaked at $(b^{*},\varphi^{*})$ such that
\begin{equation}\label{eq:cla-sca}
    \theta+\frac{2}{|\vec{p}|}\,\frac{\d I}{\d b}(b^*)-\pi=0\,, \qquad \varphi^{*}=0 \,.
\end{equation}
Finally, since the scattering angle is $\chi:=\pi-\theta$, we can use \eqref{bdef} to obtain an expression for this classical observable as a function of the radial action to all orders:
\begin{equation}\label{eq:rad-clas}
    \chi=2\,\frac{\d I_{\ell^{*}}(r=\infty)}{\d\ell}\,,
\end{equation}
which is still valid for large angles\footnote{Of course, this relation can also be derived using standard Hamilton-Jacobi analysis~\cite{Landau:1976}; what is relevant for us is that it can be related to scattering amplitudes.}.

%For small angles, one may replace $\theta$ with $\sin\theta$, thus eliminating the prefactor in \eqref{eq:2spl}. The same substitution can be made in the exponent, allowing $|\vec{p}|\,\theta\,b\,\cos\varphi$ to be interpreted as a scalar product between two vectors, such that amplitude reduces to the eikonal representation \eqref{eq:lambda-match}. However, for arbitrary angles the exponent cannot be interpreted as a scalar product between vectors due to the linearity of $\theta$. Nevertheless, the amplitude \eqref{eq:2spl} is valid for \emph{all} angles, and its saddle point approximation gives \eqref{eq:cla-sca}. This holds to all orders, without any small angle approximations.

%%%%%%%%%%%%%%%%%%%%%%%%%%
%%%%%%%%%%%%%%%%%%%%%%%%%%

\section{Photon emission}\label{sec:photons}
At 2-points, we have seen that the WKB approximation for the external states captures the full elastic scattering amplitude through the leading-order correction to the HPF in the weak field expansion. In particular, the fact that the 2-point amplitude is localized as an asymptotic boundary term means that all contributions come from the purely radial part of $S^{(1)}$, which is controlled by the radial action of the background. Beyond elastic scattering, when the emission of radiation plays a role, this is no longer the case. 

In this section, we compute the semiclassical amplitude for photon emission from a charged scalar scattering on the Coulomb background. After showing that this amplitude is controlled by the HPF, we consider the classical weak-field limit of the amplitude in the Coulomb background, recovering the classical part of the probe limit of 5-point scattering between two charged scalars with single photon emission in a trivial vacuum. This confirms that our semiclassical amplitude contains the expected physical information at leading order in perturbation theory, but also demonstrates that the radial action is not sufficient to describe classical two-body physics in the presence of emitted radiation at infinity. Indeed, the full angular dependence of the HPF is required to obtain the correct perturbative result.

%%%%%%%%%%%%%%%%%%%%%%%%%%%%%

\subsection{Semiclassical photon emission amplitude}

Following the perturbiner description of tree-level scattering amplitudes (cf., \cite{DeWitt:1967ub,Arefeva:1974jv,Abbott:1983zw,Jevicki:1987ax,Rosly:1996cp,Rosly:1996vr,Rosly:1997ap,Selivanov:1999as,Adamo:2017nia,Adamo:2020qru,Adamo:2021rfq}), the 3-point amplitude for photon emission from a complex, charged scalar in a background gauge field is given by ($\im$ times) the tri-linear terms of the scalar QED action, evaluated on solutions of the free, background-coupled equations of motion: 
\be\label{3pho1}
\bra{p',k}\mathcal{S}\ket{p} = -e\,\int\d^{4}x\left(a_{k\,\mu}^{\mathrm{out}}\,\bar{\phi}_{p}^{\mathrm{in}}\,D^{\mu}\phi_{p'}^{\mathrm{out}}-a_{k\,\mu}^{\mathrm{out}}\,({\overline D}^{\mu}\bar{\phi}_{p}^{\mathrm{in}})\,\phi_{p'}^{\mathrm{out}}\right)\,,
\ee
where $\phi_{p}^{\mathrm{in}}$, $\phi_{p'}^{\mathrm{out}}$ are the incoming and outgoing scalar fields with momenta $p$ and $p'$, respectively $a_{k\,\mu}^{\mathrm{out}}$ is the outgoing photon with (massless) momentum $k$, and $D_{\mu}=\partial_\mu-\im e A_{\mu}$ is the covariant derivative defined by the background gauge field. To obtain the \emph{semiclassical} scattering amplitude, we simply evaluate this expression using wavefunctions defined by the WKB approximation -- and hence the HPF of the Coulomb background -- in accordance with Definition~\ref{def:sca}. 

Since the photon is massless and does not interact with the background gauge field, it can be described exactly by an ordinary plane wave momentum eigenstate, while the scalar wavefunctions are given by the general solutions \eqref{gensolfin}\footnote{As amplitudes at 3-points and higher are not localized to a spacetime boundary, both incoming and outgoing wavefunctions must be fully dressed by the background.}. Explicitly, we have:
\be
\begin{split}
 \bar{\phi}_{p}^{\mathrm{in}} &=\int \d\Phi(l)\,\overline{\Lambda^{p}(l)}\,\e^{-\im\,S_{l}}\Big|_{l^0=p^0}\,, \\
 \phi_{p'}^{\mathrm{out}} &=\int\d\Phi(l')\,\Lambda^{p'}(l')\,\e^{\im\,S_{l'}}\Big|_{(l')^0=(p')^0}\,, \\
 a_{k\,\mu}^{\mathrm{out}} &=\varepsilon_{\mu}\,\e^{\im\,k\cdot x}\,,
\end{split}
\ee
where $\varepsilon_{\mu}$ is the photon polarization vector, on-shell with respect to the photon momentum $k^{\mu}$ in Lorenz gauge (i.e., $k^2=0=k\cdot\varepsilon$). the massless photon momentum. With these external wavefunctions, the semiclassical photon emission amplitude is
\begin{multline}\label{3-point-terms}
\bra{p',k}\mathcal{S}\ket{p}=\im\,e\,\int\d^{4}x\,\d\Phi(l)\,\d\Phi(l')\,\overline{\Lambda^{p}(l)}\,\Lambda^{p'}(l') \\ 
   \times \varepsilon\cdot\left(\partial S_{l}+\partial S_{l'}+2\,e\,A\right)\,\e^{\im\,(k\cdot x+S_{l'}-S_{l})}\,\Big|^{l^0=p^0}_{(l')^0=(p')^0}\,,
\end{multline}
with $A_\mu$ given explicitly by \eqref{CoulSchw}. Aside from performing the trivial time integral (which results in an energy-conserving delta function), this is as far as the semiclassical scattering amplitude can be evaluated analytically (at least, without making further approximations or simplifications).

This complexity is simply an example of the more general fact that amplitudes on strong backgrounds (even with WKB-exact wavefunctions) are generically highly non-trivial functions of the scattering data (cf., \cite{DiPiazza:2011tq,Gonoskov:2021hwf,Fedotov:2022ely}). Unlike in a trivial vacuum, it is not usually possible to perform all spacetime vertex integrals, even at low numbers of points.

Given this complexity, it is natural to ask if there are any checks that we can perform on our results. Clearly, the semiclassical amplitude \eqref{3-point-terms} contains \emph{some} of the information in the full tree-level 3-point amplitude for photon emission on the Coulomb background, but it is not at all obvious that this corresponds to the probe limit of leading classical, perturbative contributions to photon emission. To check this, one must re-expand \eqref{3-point-terms} in powers of the coupling to the background to see if an appropriate, purely perturbative result is obtained.

%%%%%%%%%%%%%%%%%%%%%%%%%%%%

\subsection{The classical and weak field limits}

If the semiclassical 3-point amplitude is truly capturing classical perturbative physics, then the leading classical contribution to $\bra{p',k}\mathcal{S}\ket{p}$ in an expansion in the coupling to the Coulomb background $Q$ should recover part of the 5-point tree-level perturbative amplitude for two massive scalar charges to scatter and emit a photon. The part of this amplitude we should recover is its leading classical behaviour (since we work in a WKB expansion) in the \emph{probe limit}, in which the recoil of one charge (which is essentially a background, since it is not affected by the scattering) is neglected.

The expansion of \eqref{3-point-terms} to first order in $Q$ contains two qualitatively distinct contributions. One set of terms will arise by taking the order $Q$ contributions from each factor of the matching coefficients $\overline{\Lambda^p(l)}$ and $\Lambda^{p'}(l')$, with all occurrences of the HPF being restricted to $S^{(0)}$. In other words, taking factors of $Q$ from the first line of \eqref{3-point-terms} only. The other set of terms arises from taking order $Q$ contributions only from the second line of \eqref{3-point-terms}, with only order $Q^0$ contributions from the matching coefficients. 

Let us begin by considering the first set of contributions to the weak-field limit:
\begin{multline}\label{4pt1}
e\,\int\d^{4}x\,\d^{2}\Omega_l\,\d^2\Omega_{l'}\,\left(\overline{f^{p}(|\vec{p}|,\,\hat{l})}\,f^{p'}(|\vec{p}\,^{\prime}|,\,\hat{l}\,^{\prime})\right)\Big|_{O(Q)} \,\varepsilon\cdot(l+l')\,\e^{\im\,(k+l'-l)\cdot x}\,\Big|^{l^0=p^0}_{(l')^0=(p')^0} \\
= 2e\,\int\hat{\delta}^{4}(k+l'-l)\,\d^{2}\Omega_l\,\d^2\Omega_{l'}\,\left(\overline{f^{p}(|\vec{p}|,\,\hat{l})}\,f^{p'}(|\vec{p}\,^{\prime}|,\,\hat{l}\,^{\prime})\right)\Big|_{O(Q)} \,\varepsilon\cdot l\,\Big|^{l^0=p^0}_{(l')^0=(p')^0}\,,
\end{multline}
ignoring irrelevant overall factors. As this is proportional to on-shell 3-point momentum conserving delta functions, it is immediately vanishing; however, it is interesting to see that these contributions also vanish if one first takes the classical limit rather than using the overall momentum conserving delta functions.

As we saw in Section~\ref{sec:2pt}, in the classical limit the elastic amplitude behaves as an eikonal-like integral, meaning the weak field expansion has the form
\be\label{elampexpand}
f^{p}(|\vec{p}|,\,\hat{l})=\delta^{2}_{\Omega_{l}}(\hat{p}-\hat{l})+\im\,\cA_{4}(\hat{l}-\hat{p}) + O(Q^2)\,,
\ee
where $\cA_4$ is the tree-level, single photon exchange amplitude between two charged scalars with exchanged momentum $\hat{l}-\hat{p}$. This arises through the eikonal phase, which is precisely the inverse Fourier transform of $\cA_4$. Furthermore, at leading order in the classical limit, the massless photon momentum $k$ scales as $\hbar$ times the classical wavenumber (cf., \cite{Kosower:2018adc}). Thus, in the classical, weak field limit the contribution \eqref{4pt1} is given by 
\begin{align}
\im\,e^2\,Q\!\int\hat{\delta}^{4}(l'-l)\,\d^{2}\Omega_l\, &\d^2\Omega_{l'}\,\varepsilon\cdot l \left(\delta^2_{\Omega_{l}}(\hat{p}-\hat{l})\,\cA_{4}(\hat{l}\,^{\prime}-\hat{p}\,^{\prime})-\delta^2_{\Omega_{l'}}(\hat{p}\,^{\prime}-\hat{l}\,^{\prime})\,\overline{\cA_{4}}(\hat{p}-\hat{l})\right)\Big|^{l^0=p^0}_{(l')^0=(p')^0} \nonumber \\
&=\im\,e^{2}\,Q\,\varepsilon\cdot p\,\left(\cA_{4}(\hat{p}-\hat{p}\,^{\prime})-\overline{\cA_{4}}(\hat{p}-\hat{p}\,^{\prime})\right) \nonumber \\
&=-2\,e^2\,Q\,\varepsilon\cdot p\,\mathrm{Im}\,\cA_{4}(\hat{p}-\hat{p}\,^{\prime})=0\,. \label{4pt2}
\end{align}
In the second and third lines, irrelevant energy conserving delta functions have been omitted, and the imaginary part of $\cA_4$ vanishes as it is a real function of the Mandelstam invariants (i.e., $s/t$), or equivalently as a consequence of the optical theorem.

\medskip
In other words, the contributions to the classical, weak field limit of $\bra{p',k}\mathcal{S}\ket{p}$ which arise from perturbatively expanding the matching coefficients vanish (which should be contrasted with the case of the $2$-point amplitude), and the \emph{only} non-trivial contributions arise from the perturbative expansion of the HPF or the explicit background insertion in \eqref{3-point-terms}. Collecting these terms requires expanding the WKB exponents, so we now replace $S_p(x) \to p\cdot x + S^{(1)}_p$ as in (\ref{S1em}), with an additional subscript to keep track of the scattering momenta, and keep terms linear in $Q$. With this, (\ref{3-point-terms}) simplifies to
\be\label{3-point-terms-2}
\begin{split}
\bra{p',k}\mathcal{S}\ket{p}\big|_{Q} = \im e \int\!\ud^4x\, \e^{\im\, (k+p'-p)\cdot x}\,\bigg[
    &\varepsilon \cdot (p+p') \big(\im \,S^{(1)}_{p'}(x) - \im\, S^{(1)}_p(x)\big) \\
    &+ \varepsilon\cdot \partial \big(S^{(1)}_{p'}(x) + S^{(1)}_{p}(x)\big)
    + 2 e\, \varepsilon\cdot A(x)\bigg]\,,
\end{split}
\ee
in which the dressing factors have been replaced by their zeroth-order expressions, given by the first term in \eqref{elampexpand}, allowing the integrals over $l$ and $l'$ to be performed.

The position integrals can now be performed as a simple Fourier transform (as taking the classical and weak field limits removes all non-trivial $x$--dependence from the exponent). The Fourier transform of $S_p^{(1)}$ is
\be\label{Fourier-conv-1}
    S^{(1)}_p(x) := -\im \int\!{\hat \ud}^4\ell \,
    \e^{-\im \ell \cdot x}\, \tilde{S}(p;\ell)\;,
    \qquad
    \tilde{S}(p;\ell)= e\,Q\, \frac{\hat\delta(U\cdot \ell)}{|\vec{\ell}|^2}\, \frac{U\cdot p}{\ell\cdot p} \;,
\ee
where the factor of $-\im$ is included for convenience, while the Fourier transform of the background Coulomb field is
\be\label{Fourier-conv-2}
     {\tilde A}_\mu(\ell) = U_\mu\,\frac{Q\, \hat\delta(U\cdot \ell)}{|\vec{\ell}|^2} \;.
\ee
With this, (\ref{3-point-terms-2}) becomes
\be
\bra{p',k}\mathcal{S}\ket{p}\big|_{Q} =\im e 
    \bigg[
    \varepsilon\cdot (p+p')(\tilde{S}(p';\q) - \tilde{S}(p;\q))
    - \varepsilon\cdot \q\, (\tilde{S}(p';\q) + \tilde{S}(p;\q))
    + 2 e\, \varepsilon\cdot {\tilde A}(\q)\bigg] \;, \label{eq:HPF-amp}
\ee
in which $\q = k + p'-p$ is the total momentum transfer. It is now apparent that the leading classical and weak field limit of our $3$-point amplitude is fully determined by the Fourier transform of the leading WKB phase $S^{(1)}$.

At this point, to obtain the \emph{classical}, weak field limit of $\bra{p',k}\mathcal{S}\ket{p}$ we need to work consistently only to leading order in the classical limit. Recall that the massless momentum $k$ scales as $\hbar$ times its classical wavenumber, and we also expect the recoil of the massive scalar probe to be small compared to its own rest mass in the classical limit. We thus write $p'_\mu=p_\mu + \Q_\mu$ in which $\Q$ scales as $\hbar$ to leading order~\cite{Kosower:2018adc}. In terms of these variables
\be\label{eq:Fourier-last}
    \bra{p',k}\mathcal{S}\ket{p}\big|_{Q} = 2\im\, e\, 
    \bigg[
    \varepsilon\cdot p\,\big(\tilde{S}(p+\Q;\q) - \tilde{S}(p;\q)\big)
    - \varepsilon\cdot \Q \, \tilde{S}(p;\q)
    + e\, \varepsilon\cdot {\tilde A}(\q)\bigg] \;.
\ee
Taking the classical limit of (\ref{eq:Fourier-last}) is then equivalent to extracting the leading term in a Taylor expansion in which both $k$ and $q$ are of the same small order (i.e., $k,q\sim\hbar$). When performing this expansion it is useful to note that, by definition, $\Q^2 +2 p\cdot \Q = 0$ and so $p\cdot \Q$ is actually of order $\hbar^2$. 

Let us consider this expansion for each of the various terms in \eqref{eq:Fourier-last} explicitly. The entire expression is proportional to an overall factor of $eQ \hat\delta(U\cdot \q)/|\vec{\q}|^2$, so we begin by  stripping this off from each term and then expanding. For instance, 
\be\label{eq:diff-A1}
    \varepsilon\cdot \Q \, \tilde{S}(p;\q) \propto \varepsilon\cdot \Q\, \frac{U\cdot p}{(k+\Q)\cdot p} \xrightarrow{\hbar\to0} \varepsilon\cdot \Q \,\frac{U\cdot p}{k\cdot p}+O(\hbar) \,,
\ee
where the leading term is $\hbar$-independent. The expansion of the combination $\tilde{S}(p+\Q;\q) - \tilde{S}(p;\q)$ is slightly more subtle:
\begin{equation}\label{eq:diff-A2}
\begin{split}
    \tilde{S}(p+\Q;\q) - \tilde{S}(p;\q) &\propto \frac{U\cdot (p+\Q)}{(k+\Q)\cdot(p+\Q)} - \frac{U\cdot p}{(k+\Q)\cdot p} \\
    & = U\cdot p\, \bigg[\frac{1}{(k+\Q)\cdot (p+\Q)} - \frac{1}{(k+\Q)\cdot p}\bigg] + \frac{U\cdot \Q}{(k+\Q)\cdot(p+\Q)} \;.
\end{split}
\end{equation}
Each of the terms in the large brackets here appears to have super-classical ($\sim\hbar^{-1}$) behaviour, but this cancels between them leaving $\hbar$-independent leading behaviour:  
\begin{equation}\label{eq:diff-A3}
\begin{split}
    \frac{1}{(k+\Q)\cdot (p+\Q)} - \frac{1}{(k+\Q)\cdot p} = \cancel{{\frac{1}{k\cdot p}} - \frac{1}{k\cdot p}} - \frac{k\cdot \Q + \Q^2}{(k\cdot p)^2} \xrightarrow{\hbar\to0} \frac{k\cdot \Q}{(k\cdot p)^2}+O(\hbar) \;,
\end{split}
\end{equation}
so that 
\be\label{eq:diff-A4}
\tilde{S}(p+\Q;\q) - \tilde{S}(p;\q) \propto U\cdot p\,\frac{k\cdot\Q}{(k\cdot p)^2}+\frac{U\cdot q}{k\cdot p}+O(\hbar)\,,
\ee
as desired. Here, we imposed that the total recoil $k+q$ is small  (to stay within the regime of validity for the background field approach), which means we only keep the leading term in $1/(k+q)^2$ and, in effect, allows us to use the identity $\Q^2+2k\cdot \Q= 0$ in the numerator. We are thus working to leading order in $\rho:=|q+k|/M$, which is the recoil of the heavy particle in units of its mass; we denote this by `LO($\rho$)'.

Assembling (\ref{eq:diff-A1})--(\ref{eq:diff-A4}), the final result for the leading classical and weak field limit of the $3$-point semiclassical amplitude on a Coulomb field is thus   
\be\label{eq:WKB-3point-QED-pert2}
\begin{split}
   \lim_{\hbar \to 0} \bra{p',k}\mathcal{S}\ket{p}\big|^{\text{LO}(\rho)}_{e^2 Q} =
    &{\hat \delta}\big(U\cdot(\Q+ k)\big)
    \frac{2\im e^2 Q}{(k+\Q)^2} \times \\
    &\bigg[
    - \varepsilon\cdot U
    + \frac{\varepsilon\cdot \Q}{k\cdot p} U\cdot p
    - \frac{\varepsilon\cdot p}{k\cdot p} U\cdot \Q - \frac{\varepsilon\cdot p\, U\cdot p\, k\cdot\Q}{(k\cdot p)^2}
    \bigg] \;.
\end{split}
\ee 
This result can now be compared to direct calculations in perturbative scalar QED; the necessary results are provided in Appendix~\ref{app:5point}. Let $\mathcal{A}_5$ be the perturbative $5$-point amplitude for photon emission (momentum $k$) in the scattering of two massive scalar charges ($p\to p'$ and $P=M U \to P'$), stripped of its momentum-conserving delta functions. We find that our $3$-point semiclassical photon emission amplitude on the Coulomb background is directly related to $\mathcal{A}_5$ via
\begin{equation}
\label{eq:HPF-amp2}
    \lim_{\hbar \to 0} \,
    \bra{p',k}\mathcal{S}\ket{p}\big|^{\text{LO}(\rho)}_{e^2 Q} = \lim_{\hbar \to 0} \, \frac{\hat{\delta}(U \cdot (q+k))}{2M}  \: \mathcal{A}_{5}^{\text{LO}(\rho)}(p,P \rightarrow p+q,P-q-k,k) \;.
\end{equation}
This relationship is sufficient to guarantee that our semiclassical amplitude will reproduce known classical observables, such as the waveform~\cite{Cristofoli:2021vyo}, in the weak field and probe limits. A further check is provided by the literature: (\ref{eq:WKB-3point-QED-pert2}) recovers the extreme mass limit ratio of e.g.~(5.48) in~\cite{Kosower:2018adc}, which is itself the leading classical limit of the full five-point amplitude. As an aside, this demonstrates the commutativity of the probe and classical limits. 

%%%%%%%%%%%%%%%%%%%%%%%%%%
%%%%%%%%%%%%%%%%%%%%%%%%%%
\subsection{The 5-point amplitude and Hamilton's principal function}
If the Hamilton-Jacobi equation is separable, the solution of the radial part is the radial action of the system, $I(r)$. It is often assumed that the radial action alone is sufficient to describe a classical two-body system (at least in the probe limit). However, as discussed in Section~\ref{sec:all-from-radial}, the results here show that it will \emph{not} hold when, for instance, considering the classical waveform at infinity.

Working to leading order in the self-force expansion, radiative observables are controlled by the $3$-point amplitude (\ref{3-point-terms}). 
Investigating even the simplest weak-field limit of this amplitude, we have now seen in (\ref{eq:HPF-amp}) and (\ref{eq:HPF-amp2}) that it is determined by the Fourier transform of the leading order HPF $S^{(1)}(x)$, which comprises both the radial action and non-trivial angular dependence, as is evident from the explicit position-space representation (\ref{S1em}). Had we not used the full HPF, but only the radial action, we would not have recovered
the correct $5$-point amplitude.

This is perhaps the simplest counterexample to the claim that the radial action controls all dynamics of a point particle on a (spherically symmetric) background.

We conclude this section by rewriting the result (\ref{eq:HPF-amp2}) in a manner which echoes the known relation between the radial action and the perturbative $4$-point amplitude. Inspecting \eqref{eq:WKB-3point-QED-pert2}, we observe that it can be written as
\be\label{eq:WKB-reprise}
\begin{split}
    \lim_{\hbar \to 0} \,
    \bra{p',k}\mathcal{S}\ket{p}\big|^{\text{LO}(\rho)}_{e^2 Q} = {\hat \delta}\big(U\cdot \ell\big)
    \frac{2\im e^2 Q}{\ell^2}\,
    \bigg[
    \frac{\varepsilon\cdot \ell}{\ell\cdot p}\, U\cdot p
    + \frac{\varepsilon\cdot p}{\ell\cdot p}\, U\cdot k - \frac{\varepsilon\cdot p\, U\cdot p\, k\cdot\ell}{(\ell\cdot p)^2}
    \bigg] \;,
\end{split}
\ee
in which $\ell\equiv q + k$, we have used $q\cdot p=0$ and, for simplicity, chosen the gauge such that $U\cdot\varepsilon =0$. We have also used the delta function to flip the sign of the second term, because doing so makes it clear that the entire result can be re-packaged in terms of the Fourier transform of the HPF (\ref{Fourier-conv-1}):
\be\label{eq:WKB-reprise2}
\begin{split}
      \lim_{\hbar \to 0} \,
    \bra{p',k}\mathcal{S}\ket{p}\big|^{\text{LO}(\rho)}_{e^2 Q} &= {\hat \delta}\big(U\cdot \ell\big)\,
    \frac{2\im e^2 Q}{\ell^2}\,
    \bigg[
    \frac{\varepsilon\cdot \ell}{\ell\cdot p}\, U\cdot p
    + \varepsilon\cdot p \, k \cdot \partial_p\left( \frac{U\cdot p}{\ell\cdot p}\right) \bigg] \\
    & = - 2e\, 
    \Big[
    \varepsilon\cdot \ell \,  {\tilde S}(p;\ell)
    + \varepsilon\cdot p\, k \cdot \partial_p \, {\tilde S}(p;\ell) \Big] \;.
\end{split}
\ee
This Fourier transform can be undone by integrating over $\ell$, keeping in mind that the polarisation vectors depend on $k=\ell-q \to -\im \partial_x - q$. We could either express the inverse Fourier transform of (\ref{eq:WKB-reprise2}) in terms of $\hat\varepsilon :=\varepsilon(-\im\partial_x-q)$ or, since we have fixed the gauge, simply strip the polarisation vectors from both sides of (\ref{eq:WKB-reprise2}) before performing the transform. In this case we find, expressing the classical limit of $
    \bra{p',k}\mathcal{S}\ket{p}\big|^{\text{LO}(\rho)}_{e^2 Q}$ in terms of $\mathcal{A}_5\equiv \varepsilon^\mu \mathcal{A}_{5\mu}$ from (\ref{eq:HPF-amp2}),
\be\label{eq:WKB-reprise3}
\begin{split}
   \lim_{\hbar \to 0} \int\!\ud^4\ell\, \e^{-\im\ell\cdot x} \, \frac{\hat{\delta}(U \cdot \ell)}{2M}  \: \mathcal{A}_{5\mu}^{\text{LO}(\rho)} &(p,P \rightarrow p+q,P-\ell,\ell-q)  \\
   &= 2e\left[\frac{\partial S_p^{(1)}}{\partial x^{\mu}}(x)  -\im p_\mu\, k^{\nu}\,\frac{\partial S_p^{(1)}}{\partial p^{\nu}}(x)\right] \;.
\end{split}
\ee
Thus, the Fourier transform of the $5$-point amplitude is a linear function of the HPF and its derivatives (with respect to both position and asymptotic momentum). This is highly reminiscent of the result that the Fourier transform of the $4$-point amplitude is the leading order radial action, but again emphasises that it is the HPF that controls radiation.
%

%%%%%%%%%%%%%%%%%%%%%%%%%%%
%%%%%%%%%%%%%%%%%%%%%%%%%%%
\section{Graviton emission}\label{sec:gravitons}

We now turn to the semiclassical graviton emission amplitude on a Schwarzschild spacetime. This computation is significantly more complicated than its scalar QED counterpart for two reasons: firstly, the emitted graviton `sees' the background field (unlike the photon in QED) so defining the semiclassical graviton wavefunction is non-trivial; and secondly, because effects due to the event horizon must be accounted for in the fully non-linear black hole background. At 2-points, the issue of the event horizon could be ignored with the physically reasonable assumption that elastic scattering occurs at sufficiently large impact parameter, or equivalently that the 2-point amplitude only receives contributions from the asymptotic boundary. However, at 3-points and beyond, interactions must be integrated over the whole spacetime manifold and the event horizon simply cannot be ignored.

In this section, we outline how the semiclassical graviton wavefunction on fully non-linear Schwarzschild could be determined within our framework, but to circumvent dealing with horizon effects -- and to present a more concrete calculation -- we then simplify to a \emph{linearised} Schwarzschild background, where the event horizon does not play a role. Even with this simplification, determining the graviton wavefunction and computing the 3-point semiclassical emission amplitude is non-trivial, and we confirm that this is controlled by the HPF of the background. As in the QED case, we show that the classical weak-field limit of the amplitude on linearised Schwarzschild recovers the classical part of the probe limit of 5-point scattering between massive scalars with single graviton emission in Minkowski spacetime. 
%%%%%%%%%%%%%%%%%%%%%%%%%

\subsection{Semiclassical graviton states}

We begin by considering a generic vacuum spacetime background with metric $g_{\mu\nu}$, and a linearised metric perturbation $h_{\mu\nu}$ on this background. Define 
\be\label{tsgrav}
\bar{h}_{\mu\nu}:=h_{\mu\nu}-\frac{1}{2}\,g_{\mu\nu}\,h^{\sigma}{}_{\sigma}\,,
\ee
and impose the covariant Lorenz gauge
\be\label{colorg}
\nabla^{\mu}\bar{h}_{\mu\nu}=0\,,
\ee
where all indices are raised and lowered with background metric $g$ and $\nabla_{\mu}$ is the Levi-Civita connection of the background. In this gauge, the linearised Einstein equations governing the gravitational perturbation become:
\be\label{lineEFEs}
\nabla^2 \bar{h}_{\mu\nu}+2\,R_{\mu\rho\nu\sigma}\,\bar{h}^{\rho\sigma}=0\,,
\ee
where $\nabla^2:=g^{\alpha\beta}\,\nabla_{\alpha}\nabla_{\beta}$, $R_{\mu\rho\nu\sigma}$ is the Riemann curvature tensor of the background and we have used the assumption that the background is vacuum, and hence that its Ricci tensor vanishes. 

There is no $\hbar$ entering this equation: the graviton is a massless wave and so the linearised Einstein equation is classically exact. However, it is nevertheless useful to make an ansatz for the graviton wavefunction which is motivated by the desire to have a Fourier basis:
\be\label{gravWKB1}
\bar{h}_{\mu\nu}(x)=\cE_{\mu\nu}(x)\,\e^{\im\,S(x)}\,,
\ee
where $S$ is the HPF solving the \emph{massless} Hamilton-Jacobi equation on the background
\be\label{masslessHPF}
S(x)=k\cdot x+\frac{G\,\cP^{\mu\nu}\,k_{\mu}\,k_{\nu}}{|\vec{k}|}\,\log(|\vec{k}|r+\vec{k}\cdot\vec{r})+O(G^2)\,.
\ee
Note that the graviton polarization tensor, $\cE_{\mu\nu}(x)$, acquires non-trivial spacetime dependence due to its interaction with the background. The choice to factor out the HPF phase is motivated by the form of known exact graviton wavefunctions in other background fields, for example plane waves, see~\cite{Adamo:2017nia}. As we will see, the choice is natural: the asymptotic behavior at spatial infinity of our graviton state contains an accumulated phase similar to that in the scalar wavefunction. This is completely consistent, since it matches the asymptotic behavior of a gravitational perturbation around Schwarzschild, and it is equivalent to requiring, as a boundary condition, a free phase in tortoise coordinates (see, for example, Section 12.2.5 of \cite{Maggiore:2018sht}). In particular, the polarisation tensor becomes equal to the free-field polarisation at large distance, simplifying our matching conditions.

A straightforward calculation with this ansatz shows that the linearised Einstein equations and Lorenz gauge condition become
\be\label{lEFE3}
\nabla^2\cE_{\mu\nu}+\im\,\cE_{\mu\nu}\,\nabla^{2}S+2\im\,\partial_{\alpha}S\,\nabla^{\alpha}\cE_{\mu\nu}+2\,R_{\mu\rho\nu\sigma}\,\cE^{\rho\sigma}=0\,,
\ee
and
\be\label{lLG}
\nabla^{\mu}\cE_{\mu\nu}+\im\,\cE_{\mu\nu}\,\partial^{\mu}S=0\,,
\ee
respectively. In particular, this means that with the ansatz \eqref{gravWKB1}, the background-dressed semiclassical graviton polarization $\cE_{\mu\nu}$ is determined by the HPF (and the background geometry) through \eqref{lEFE3} and gauge consistency condition \eqref{lLG}\footnote{Observe that, as the graviton wavefunction is meant to describe a classical wave, its background-dressed polarization does not obey the transport equation and the gauge consistency condition of the strict geometric optics limit (for more on this limit, see~\cite{Maggiore:2007ulw,DeWitt:2011nnj,Harte:2018wni,Harte:2019tid}).}. Note that if we additionally impose traceless gauge (i.e., $h^{\mu}_{\mu}=0$), then $\bar{h}_{\mu\nu}=h_{\mu\nu}$ and these are the equations for the graviton polarization itself.

We make the additional assumption that the dressed polarization admits a weak field expansion
\be\label{polexpand}
\cE_{\mu\nu}(x)=\varepsilon_{\mu\nu}+\sum_{n=1}^{\infty}\cE_{\mu\nu}^{(n)}(x)\,,
\ee
with $\varepsilon_{\mu\nu}$ the on-shell graviton polarization in Minkowski spacetime in transverse traceless gauge ($\eta^{\mu\nu}\varepsilon_{\mu\nu}=0=\eta^{\sigma\mu}\,k_{\sigma}\,\varepsilon_{\mu\nu}$) and $\cE_{\mu\nu}^{(n)}$ of order $G^n$. Thus, the dressed polarization tensor can be determined order-by-order by solving the linearised Einstein equation with the weak field expansion of the HPF itself. For instance, using the weak field expansion \eqref{BFexp} of the background spacetime metric, \eqref{lEFE3} at linear order in $G$ becomes:
\be\label{wlEFE1}
\left(\Box+2\im\,k\cdot\partial\right)\cE^{(1)}_{\mu\nu}+\im\,\varepsilon_{\mu\nu}\left(\Box S^{(1)}-k_{\sigma}\,\eta^{\alpha\beta}\,\Gamma^{(1)\,\sigma}_{\alpha\beta}\right)-4\im\,k^{\alpha}\,\varepsilon_{\sigma(\mu}\,\Gamma^{(1)\,\sigma}_{\nu)\alpha}+2\,R^{(1)}_{\mu\rho\nu\sigma}\,\varepsilon^{\rho\sigma}=0\,,
\ee
where all indices are now raised and lowered with the \emph{Minkowski} metric, and $\Gamma^{(1)\,\alpha}_{\beta\gamma}$ and $R^{(1)}_{\mu\rho\nu\sigma}$ are the linearised Christoffel symbols and Riemann curvature tensor, both constructed from $H_{\mu\nu}$. 

With the specification that we are working with the Schwarzschild metric, for which $H_{\mu\nu}$ is given by \eqref{CoulSchw}, it follows that we can solve for $\cE^{(1)}_{\mu\nu}$ using a Green's function:
\begin{multline}\label{wdPol1}
\cE_{\mu \nu}^{(1)}(x) =  -  2\, \varepsilon_{\beta(\mu}\,k^{\alpha}\int \hat{\d}^4\ell \:\frac{\e^{-\im\,\ell\cdot x}}{\ell^2 - 2 \ell \cdot k+\im\, \epsilon}\, \bigg(\tilde{H}_{\nu)}^{\beta}\,\ell_{\alpha}+\ell_{\nu)}\,\tilde{H}_{\alpha}^{\beta}-\tilde{H}_{\nu)\alpha}\,\ell^{\beta} \bigg)\\- \im\,
    \varepsilon_{\mu \nu} \int \hat{\d}^4\ell \:\frac{|\vec{\ell}|^2\,\e^{- \im\, \ell\cdot x}}{\ell^2 - 2 \ell \cdot k+\im\, \epsilon}\, \tilde{S}(k;\ell) \\
   - \varepsilon^{\rho \sigma} \int \hat{\d}^4 \ell\, \frac{\e^{-\im\, \ell \cdot x}}{\ell^2 - 2 \ell \cdot k + \im\,\epsilon} \left( - \ell_{\mu} \ell_{\nu}\, \tilde{H}_{\rho \sigma} + \ell_{\nu} \ell_{\rho}\, \tilde{H}_{\mu \sigma} + \ell_{\sigma} \ell_{\mu}\, \tilde{H}_{\rho \nu} - \ell_{\sigma} \ell_{\rho}\, \tilde{H}_{\mu \nu} \right)\,,
\end{multline} 
where $\im \,\epsilon$ denotes a choice of contour prescription for inverting the differential operator acting on $\cE^{(1)}_{\mu\nu}$ in \eqref{wlEFE1} in momentum space. In this expression
\be\label{SchwFTs}
\tilde{H}_{\mu\nu}=4\pi\,G\,\frac{\cP_{\mu\nu}\,\hat{\delta}(U\cdot \ell)}{|\vec{\ell}|^2}\,, \qquad \tilde{S}(k;\ell)=2\pi\im\,G\,\frac{\hat{\delta}(U\cdot \ell)\,\cP^{\alpha\beta}\,k_{\alpha}\,k_{\beta}}{|\vec{\ell}|^2\,  \ell\cdot k}\,,
\ee
 so it follows that all of the $\ell_{\mu}$ appearing in \eqref{wdPol1} are effectively $\vec{\ell}_\mu$ and purely spatial. Proceeding in this fashion, one can recursively solve for the dressed polarization order-by-order in the weak field expansion, similarly to how one constructs the HPF itself. It can be verified that $\cE^{(1)}_{\mu\nu}$ obeys the Lorenz gauge condition \eqref{lLG} to linear order in the coupling at leading order in $r^{-1}$, which is the regime of validity for our approximation, to this order. One can also verify that it satisfies the traceless gauge condition at this order. Additionally, it is easily seen that $\lim_{r \rightarrow \infty} \cE_{\mu \nu}^{(1)} (x) = 0$, reducing the graviton polarization to the free field polarization at large $r$. 

\medskip

Just as we did with scalar states in Section~\ref{sec:2pt}, we construct a general graviton wavefunction by taking an on-shell linear combination of the solutions to the linearised Einstein equations in the form of our ansatz:
\be\label{gravgen}
h_{\mu\nu}(x)=\int\d\Phi(k)\,\Lambda_{\mu\nu}{}^{\rho\sigma}(k)\,\cE_{\rho\sigma}(x)\,\e^{\im\,S_k(x)}\,,
\ee
where $\d\Phi(k)$ is the massless Lorentz-invariant on-shell measure and $\Lambda_{\mu\nu}{}^{\rho\sigma}(k)$ are the tensorial coefficients of the on-shell combination. Although our exposition above of the linearised Einstein equations \eqref{lEFE3} was in Lorenz gauge, this expression for a general graviton is schematically true in \emph{any} gauge: the difference with Lorenz gauge will be in the structure of the PDE determining $\cE_{\mu\nu}$ from the HPF. 

In the fully non-linear Schwarzschild spacetime, the coefficients in \eqref{gravgen} must be consistent with boundary values of exact solutions to the linearised Einstein equations both asymptotically ($r\to\infty$) \emph{and} at the event horizon ($r\to 2GM$). The mechanism for doing this is to consider the linearised Einstein equations with separation of variables; this was initially done long ago in the Regge-Wheeler gauge, where gravitational perturbations are governed by simple 1-dimensional radial Schr\"odinger equations with different potentials depending on whether they are of axial or polar type~\cite{Regge:1957td,Zerilli:1970se,Chandrasekhar:1975zza,Chandrasekhar:1985kt}. This is still the case in Lorenz gauge~\cite{Berndtson:2007gsc}, where the radial part of the perturbation is controlled by a generalized Regge-Wheeler-Zerilli equation. The presence of dissipative horizon dynamics (such as quasinormal modes) would then correspond to allowing the HPF to become \emph{complex}, with corresponding damped, non-oscillatory behaviour at the event horizon. Matching at the event horizon must also be taken into account for the scalar wavefunctions -- defined in Section~\ref{sec:2pt} only through a matching at infinity -- governed by the behaviour of the radial scalar wavefunctions of the Klein-Gordon equation in the Schwarzschild metric near the horizon (cf., \cite{Bardeen:1973xb,Rowan:1976ug,Jensen:1985in}).

While it would be extremely interesting to understand how the matching at infinity and event horizon is implemented in the fully non-linear setting, doing so explicitly is beyond the scope of this paper. To give a precise example of the general semiclassical graviton wavefunction \eqref{gravgen}, we focus on the case of the \emph{linearised} Schwarzschild metric. In this case, the expansion of the dressed polarization tensor truncates with $\cE_{\mu\nu}^{(1)}$ given by \eqref{wdPol1}. In the linearised metric, there is no event horizon and the only matching condition is at infinity, as $r\to\infty$. However, we have already seen that the dressed polarisation reduces to the free field polarization $\varepsilon_{\mu \nu}$ as $r \rightarrow \infty$. Combined with the asymptotic behaviour of the pure HPF phase part in \eqref{gravgen}, the matching coefficients automatically reduce to those of a massless scalar: 
\be\label{lingrco}
\Lambda^{k'}_{\mu\nu}{}^{\rho\sigma}(k)\Big|_{\text{lin. Schw.}}=-2\im\,\delta_{(\mu}^{\rho}\,\delta_{\nu)}^{\sigma}\,\delta(|\vec{k}|-|\vec{k}\,^{\prime}|)\,f^{k'}(|\vec{k}|,\,\hat{k})\,,
\ee
where $f^{k'}(|\vec{k}|,\,\hat{k})$ is the \emph{massless} (i.e., $m=0$) scalar elastic scattering amplitude on Schwarzschild.

In other words, the general wavefunction for a graviton of momentum $k$ in the linearised Schwarzschild metric is given by
\be\label{scgrav}
h_{k\,\mu\nu}(x)=-\im\,|\vec{k}|\int\d^{2}\Omega_{l}\,f^{k}(|\vec{k}|,\,\hat{l})\,\cE_{\mu\nu}(x)\,\e^{\im\,S_{l}(x)}\Big|_{l^0=k^0}\,,
\ee
where all 4-momenta appearing in this expression are null and $\cE_{\mu\nu}$ is implicitly on-shell with respect to $l$.

%%%%%%%%%%%%%%%%%%%%%%%%%

\subsection{Semiclassical graviton emission amplitude}

Once again following the perturbiner prescription for tree-level scattering amplitudes, the 3-point amplitude for graviton emission from a massive complex scalar in a curved background spacetime is given by the tri-linear terms in the gravitationally coupled scalar action\footnote{We work throughout with a minimally coupled scalar field. Relaxing this assumption would amount to modeling finite size effects for the object moving in the background, as is customary in an EFT language. While this does not play a role at the level of the discussion in this paper, we expect it to be necessary at higher orders to ensure finite self-force corrections to scattering observables. This is because the notion of a point particle is not assumed but derived when considering the self-force approximation. For an example of where divergences might arise, see~\cite{Barack:2023oqp}.}:
\be\label{3gra1}
\kappa\,\int\d^{4}x\,\sqrt{-|g|}\,h_{k\,\mu\nu}^{\mathrm{out}}\left[2\,\partial^{\mu}\phi_{p'}^{\mathrm{out}}\,\partial^{\nu}\bar{\phi}_{p}^{\mathrm{in}}-g^{\mu\nu}\left(\partial_{\alpha}\phi_{p'}^{\mathrm{out}}\,\partial^{\alpha}\bar{\phi}_{p}^{\mathrm{in}}-\frac{m^2}{2}\,\phi_{p'}^{\mathrm{out}}\,\bar{\phi}_{p}^{\mathrm{in}}\right)\right]\,,
\ee
where $\kappa=\sqrt{8\pi G}$ is the gravitational coupling constant, all indices are raised and lowered with the background metric $g_{\mu\nu}$ and $|g|$ is its determinant. For the fully non-linear Schwarzschild black hole, we can only give a schematic refinement of this expression, with semiclassical wavefunctions
\be
\begin{split}
 \bar{\phi}_{p}^{\mathrm{in}} &=\int \d\Phi(l)\,\overline{\Lambda^{p}(l)}\,\e^{-\im\,S_{l}}\Big|_{l^0=p^0}\,, \\
 \phi_{p'}^{\mathrm{out}} &=\int\d\Phi(l')\,\Lambda^{p'}(l')\,\e^{\im\,S_{l'}}\Big|_{(l')^0=(p')^0}\,, \\
 h_{k\,\mu\nu}^{\mathrm{out}} &=\int\d\Phi(k')\,\Lambda^{k}_{\mu\nu}{}^{\rho\sigma}(k')\,\cE_{\rho\sigma}\,\e^{\im\,S_{k'}}\Big|_{(k')^0=k^0}\,,
\end{split}
\ee
where the graviton wavefunction and dressed polarization are defined in Regge-Wheeler gauge and the scalar and tensorial matching coefficients are determined by agreement with exact solutions to the Klein-Gordon and Regge-Wheeler-Zerilli equations, respectively, at the event horizon and asymptotically. We re-emphasize that we have certainly not described this matching at any sort of technical level.
With this in mind, the semiclassical graviton emission amplitude is given schematically by:
\begin{multline}\label{3granl}
\bra{p',k}\mathcal{S}\ket{p}=-\kappa\!\int\limits_{\R^{1,3}\setminus\overline{B(r_S)}}\!\!\d^{4}x\,\sqrt{-|g|}\int \d\Phi(l)\,\d\Phi(l')\,\d\Phi(k')\,\overline{\Lambda^{p}(l)}\,\Lambda^{p'}(l')\,\Lambda^{k}_{\mu\nu}{}^{\rho\sigma}(k')\,\cE_{\rho\sigma} \\
\times\left[2\partial^{\mu}S_{l'}\,\partial^{\nu}S_l-g^{\mu\nu}\left(\partial S_{l'}\cdot\partial S_l+\frac{m^2}{2}\right)\right]\e^{\im\left(S_{k'}+S_{l'}-S_{l}\right)}\Big|^{l^0=p^0}_{(l')^0=(p')^0,\,\,(k')^0=k^0}\,,
\end{multline}
where the region of integration over spacetime is the exterior ($r>r_S=2GM$) of the black hole.

For the case of the \emph{linearised} Schwarzschild background, we can be significantly more explicit. By working in the traceless Lorenz gauge, only the first term in the integrand of \eqref{3gra1} survives, there is a well-defined S-matrix and the only matching conditions are at infinity, simplifying the structure of the outgoing graviton wavefunction \eqref{scgrav}. This leads to the semiclassical graviton emission amplitude
\begin{multline}\label{3gralin}
\bra{p',k}\mathcal{S}\ket{p}=-2\,\kappa\int\d^{4}x\,\d\Phi(l)\,\d\Phi(l')\,\d\Phi(k')\,\sqrt{-|g|}\,\overline{\Lambda^{p}(l)}\,\Lambda^{p'}(l')\,\Lambda^{k}(k') \\
\times\,\cE_{\mu\nu}\,\partial^{\mu}S_{l'}\,\partial^{\nu}S_l\,\e^{\im\left(S_{k'}+S_{l'}-S_{l}\right)}\Big|^{l^0=p^0}_{(l')^0=(p')^0,\,\,(k')^0=k^0}
\end{multline}
where $g_{\mu\nu}$ is now the linearised Schwarzschild metric in the spherical coordinates of \eqref{CoulSchw} and the HPFs are defined with respect to this metric. 

%%%%%%%%%%%%%%%%%%%%%%%%%

\subsection{The classical and weak field limits}

To check the classical, weak field limit of the semiclassical graviton emission amplitude, it suffices to start with the answer \eqref{3gralin} on linearised Schwarzschild. As in the photon emission calculation, it is convenient to divide contributions to this limit into those coming from the matching coefficients -- or equivalently, elastic amplitudes -- and everything else. In \eqref{3gralin}, there are three such matching coefficients: one for each incoming/outgoing scalar and one for the emitted graviton. It is easy to see that the contributions coming from perturbatively expanding each of these in turn to linear order in $G$ will be proportional to on-shell 3-point momentum conservation, so all such contributions vanish for exactly the same reason as in the QED calculation

Thus, the only contributions to the classical, weak field limit of $\bra{p',k}\mathcal{S}\ket{p}$ come by taking powers of $G$ from the HPF, the dressed polarization, or explicit insertions of the background metric. A straightforward calculation shows that
\begin{multline}\label{gpert1}
\bra{p',k}\mathcal{S}\ket{p}|_{\kappa^3}=-2\kappa\,\Bigg[\varepsilon_{\mu\nu}\,p^{\mu}\,p^{\nu}\left(\tilde{S}(k;k+q)+\tilde{S}(p+q;k+q)-\tilde{S}(p;k+q)\right) \\
-\varepsilon_{\mu\nu}\,p^{\mu}\,q^{\nu}\left(\tilde{S}(p;k+q)+\tilde{S}(p+q;k+q)\right)+\tilde{\cE}^{(1)}_{\mu\nu}(k+q)\,p^{\mu}\,(p+q)^{\nu} \\
+\frac{1}{2}\,\tilde{H}^{\sigma}_{\sigma}(k+q)\,\varepsilon_{\mu\nu}\,p^{\mu}\,p^{\nu}-\tilde{H}^{\mu\sigma}(k+q)\,\varepsilon_{\mu\nu}\,p^{\nu}\,(p+q)_{\sigma}-\tilde{H}^{\nu\sigma}(k+q)\,\varepsilon_{\mu\nu}\,p^{\mu}\,p_{\sigma}\Bigg]\,,
\end{multline}
with $p^{\prime}_{\mu}=p_{\mu}+q_\mu$ and the Fourier transformed quantities are defined by \eqref{SchwFTs} and \eqref{wdPol1}. To take the classical limit of this expression, we expand all quantities to leading order in the $\hbar\to0$ limit, keeping in mind that massless momenta $k_\mu$, $q_{\mu}$ scale linearly with $\hbar$ in this limit. Furthermore, as in Section~\ref{sec:photons}, we impose that the total recoil $k+q$ is small, and exploit identities such as $q^2+2p\cdot q=0$, which tell us that $p\cdot q$ scales like $\hbar^2$ in the classical limit. The result is
\begin{multline}\label{gpert2}
\lim_{\hbar\to 0} \bra{p',k}\mathcal{S}\ket{p}|^{\text{LO}(\rho)}_{\kappa^3}={\hat \delta}\big(U\cdot(q+k)\big)\,\frac{64\,M^2\,\kappa^3}{(q+k)^2}\,\varepsilon_{\mu\nu}\left[U^{\mu}\,U^{\nu}\,\frac{(k\cdot p)^2}{q^2}-2\,U^{\mu}\,q^{\nu}\,\frac{k\cdot p\,p\cdot U}{q^2}\right. \\
+\frac{q^{\mu}\,q^{\nu}}{q^2}\left((p\cdot U)^2-\frac{m^2}{2}\right)+\frac{2\,p^{\mu}\,U^{\nu}}{q^2}\left(k\cdot p\,q\cdot U-\frac{q^2}{2}\,(p\cdot U)\right)\\
-\frac{p^{\mu}\,q^{\nu}}{q^2}\left(\frac{q^2\,m^2}{2\,k\cdot p}+2\,p\cdot U\,q\cdot U-\frac{q^2\,(p\cdot U)^2}{k\cdot p}\right) \\
\left.-p^{\mu}\,p^{\nu}\left(\frac{m^2\,q^2}{8\,(k\cdot p)^2}-\frac{(p\cdot U)^2\,q^2}{4\,(k\cdot p)^2}+\frac{p\cdot U\,q\cdot U}{k\cdot p}-\frac{(q\cdot U)^2}{q^2}\right)\right]\,, 
\end{multline} 
for the classical weak field limit of the semiclassical graviton emission amplitude.

This can now be compared against the perturbative 5-point amplitude for two scalars of masses $M\gg m$ to scatter and emit a single graviton in Minkowski spacetime~\cite{Luna:2017dtq}. Let $\cA_5$ denote this tree-level amplitude, stripped of its overall momentum conserving delta functions. By comparing with the analysis of this amplitude in Appendix~\ref{app:5point}, we find the relationship
\be\label{gpert3}
\lim_{\hbar \to 0} \,
\bra{p',k}\mathcal{S}\ket{p}\big|^{\text{LO}(\rho)}_{\kappa^3} = \lim_{\hbar \to 0} \, \frac{\hat{\delta}(U \cdot (q+k))}{2M}  \: \mathcal{A}_{5}^{\text{LO}(\rho)}(p,P \rightarrow p+q,P-q-k,k)\,,
\ee
between the classical weak field limit of our semiclassical 3-point amplitude in Schwarzschild and the classical, probe limit of the perturbative 5-point amplitude in Minkowski spacetime. Once again, this demonstrates that the semiclassical amplitude will recover known classical observables in the weak field and probe limits. 

\medskip

As the Coulomb electromagnetic field and the Schwarzschild metric are related by classical double copy~\cite{Monteiro:2014cda}, one may be tempted to look for a double copy relationship between our graviton \eqref{3gralin} and photon \eqref{3-point-terms} emission amplitudes on those respective backgrounds. However, at this stage we have only superficial comments to make in this direction. Firstly, there is a fairly obvious `double copy' relationship between the first-order HPFs, replacing electromagnetic charge with a second copy of probe momentum and tensorial structure corresponding to subtracting the dilaton. Secondly, as the perturbative limits of both emission amplitudes correspond to the probe limits of the corresponding 5-point perturbative amplitudes, the known double copy relationship~\cite{Luna:2017dtq} between those amplitudes is similarly recovered in the perturbative limit.

It would, of course, be interesting to explore a more enlightening notion of double copy between the emission amplitudes that fully manifests the non-perturbative nature of the backgrounds.

%%%%%%%%%%%%%%%%%%%%%%%%%

%%%%%%%%%%%%%%%%%%%%%%%%%%%%%
\section{Conclusions}
\label{sec:concl}
Scattering amplitudes serve as the natural building blocks for
studies of the two-body problem in general relativity. When defined in a Minkowski vacuum, they provide the integrands necessary to extract classical observables within the perturbative Post-Minkowskian (PM) approximation~\cite{Kosower:2018adc}. On a generic background, they define the on-shell integrands for perturbative self-force corrections to observables of scattering orbits~\cite{Adamo:2022rmp,Adamo:2022qci}: the relevance of this perturbative scheme for the two-body problem is twofold.

First, and practically, while ground-based observatories continue to generate waveform templates by combining information from post-Newtonian calculations and numerical relativity in the strong field regime, this approach will no longer be suitable for extreme mass ratio inspiral waveforms, such as those accessible to eLISA\footnote{For more details, see~\cite{Barausse:2020rsu}, Chapter II.2 of~\cite{Barack:2018yly} and section 1.2 of \cite{Long:2022sdq}}. This motivates efforts to explore the self-force expansion using modern tools such as those coming from quantum field theory (QFT), where on-shell data defines the scattering problem. Secondly, on a conceptual level, this approach offers an intriguing way to inform other perturbative schemes, such as PM calculations. For instance, Damour demonstrated that the calculation of the gravitational scattering angle at the first self-force order determines the complete two-body potential through 4PM (i.e., order $G^4$) to all orders in the mass ratio \cite{Damour:2019lcq}, providing a concrete motivation for investigating these calculations using scattering amplitudes. This perspective opens up, in particular, possibilities to understand perturbation schemes that do not rely on weak field assumptions, using only perturbative amplitudes in vacuum and their resummation. This, in turn, would offer potential applications of powerful methods like the double copy, generalized unitarity, and BCFW recursion relations in a strong field regime.

In this work, we have already seen a few examples of this. Our main results concerning the semiclassical $2$-point (\ref{2ptc2}) and $3$-point amplitudes on Coulomb (\ref{3-point-terms}) and Schwarzschild (\ref{3gralin}) demonstrate that these on-shell quantities encode the expected weak field probe limit dynamics for the radiative sector of classical scattering. Understanding these quantities at the perturbative level solely in terms of on-shell data has allowed us to revisit the entire scattering amplitude defined on the background as a resummation of amplitudes in vacuum. This perspective reveals known relations, such as the connection between the radial action and the 4-point amplitude in vacuum~\cite{Bern:2021dqo,Kol:2021jjc}. However, for the
5-point amplitude in vacuum, the structure is more intricate. We have found that it receives contributions not only from the radial action but also from the angular part of the Hamilton's principal function. This implies that even the simplest amplitude with emission on a background carries significant information that is not available in the radial action alone.

Looking to the future, it would be interesting to explore the extraction of classical observables from the background field amplitudes that we have considered. Their use can be made systematic following~\cite{Adamo:2022rmp} and by a proper counting of the matrix elements on the background in powers of the mass ratio. For example, the first deviation from geodesic motion will appear in the form of radiation emitted by a particle moving on the background. This would result in a formula for the total power emitted, which would in turn generate a correction to geodesic motion due to momentum balance. This correction will depend on the mass of the particle. At leading order, the formula for the radiated momentum can be expressed as on-shell integral over a wavepacket $\phi(p)$ - sharply localized to ensure a well defined classical limit - and the impact parameter $b$ as:
\begin{equation}
K^{\mu}=\sum_{\eta}\int \d\Phi(p,p',l,k)\, \phi(p)\,\overline{\phi}(p')\,\e^{\im\,b\cdot (p-p')/\hbar} \bra{p'}\mathcal{S}^{\dag}\ket{l,k^{\eta}}\bra{l,k^{\eta}}\mathcal{S}\ket{p} k^{\mu} \ .
\end{equation}
Here, the sum is over the helicity of the emitted photon or graviton and the $\hbar\to0$ limit of the entire expression of the right-hand-side is implicit. 

To perform all of the integrals in this expression using our results, a combination of six eikonal-type integrals would be necessary, along with additional contributions. Although we have not considered classical observables in this paper, our approach offers an alternative method to address self-force corrections to geodesic motion, solely based on on-shell data. By contrast, the standard approach to first-order in self-force corrections relies on the `MiSaTaQuWa equations'~\cite{Mino:1996nk,Quinn:1996am}\footnote{For a perturbative treatment of the latter, see~\cite{Elkhidir:2023dco}. In this case, real and imaginary parts of loop amplitudes in vacuum play relevant but different roles in the radiative physics.}. Results for radiative observables obtained in this way (such as the radiated momentum or waveform) could be cross-checked against results found by numerically solving the Regge-Wheeler-Zerilli equations~\cite{Martel:2003jj,Martel:2005ir}. It would also be interesting to contrast the derivation of similar classical observables, which incorporate contributions of all orders in the coupling, with a conjecture presented in~\cite{Cristofoli:2021jas} regarding the final semiclassical state in a two-body scattering scenario involving radiation. When one of the objects has a significantly larger mass than the other, it would effectively acts as a background as in our calculation. Finally, it would be interesting to extend our results with radiation to a Kerr background, motivated by the possibility of describing exact geodesic motion with perturbative methods~\cite{Menezes:2022tcs,Damgaard:2022jem}. We hope to explore this in future work.

\acknowledgments

We thank Francesco Comberiati, Uri Kol, Matteo Sergola and Piotr Tourkine for conversations and useful feedback on the manuscript. AC is grateful to the Mani L. Bhaumik Institute for Theoretical Physics for their hospitality and for providing a highly stimulating environment where part of this work was developed. In particular, AC would like to thank Leor Barack, Zvi Bern, Enrico Herrmann, Julio Parra-Martinez, Donal O’Connell, Michael Ruf, Atul Sharma, Mikhail Solon and Justin Vines for useful and interesting discussions. The authors are supported by a Royal Society University Research Fellowship (TA), Leverhulme Trust grant RPG-2020-386 (TA \& AC), the STFC Consolidator grant ST/X000494/1 “Particle Physics at the Higgs Centre” (TA \& AI), and an EPSRC studentship~(SK).
\appendix

%%%%%%%%%%%%%%%%%%%%%%%%%%%%%
\section{Detailed asymptotic expansions}
\label{app:Asymp}
%%%%%%%%%%%%%%%%%%%%%%%%%%%%%

In this appendix, we provide a rigorous derivation of the matching coefficients \eqref{Lambda1}, coming from the asymptotic matching of our WKB ansatz to the known exact solution of the Klein-Gordon equation on the same backgrounds \eqref{shexp}. 

We start with the ansatz 
\begin{equation}\label{appansatz}
\phi (x) = \int \d \Phi(p)\, \Lambda(p)\, \e^{\im\, S_p (x)}\,,
\end{equation}
for some unknown coefficients $\Lambda(p)$ that will be matched onto the exact solution asymptotically. Expanding $\e^{\im\, S_p(x)}$ in spherical harmonics gives the general result
\begin{equation}
\e^{\im\, S_p (x)} = \e^{\im\, E t}\sum_{\ell = 0}^{\infty} \frac{2\ell +1}{2}\,c_\ell (r)  P_\ell (\cos \theta)\,, \label{legPolyExp}
\end{equation}
where $\cos\theta=\hat{p}\cdot\hat{x}$ and the coefficients $c_{\ell}(r)$ are determined via
\begin{equation}
c_\ell (r) = \int_0^{\pi} \d \theta\, \sin \theta\,  \e^{\im\left( S_p(r, \cos \theta) - E t\right)} P_\ell (\cos \theta)\,. \label{legTransf}
\end{equation}
The ansatz \eqref{appansatz} is thus equivalent to
\begin{equation}\label{ans2}
\phi(x) = 2 \pi  \int \d \Phi(p)\, \Lambda(p)\, \e^{\im\, Et} \sum_{\ell, m}c_\ell(r)\, Y_{\ell}^{m}(\hat{x})\, \overline{Y_{\ell}^{m}}(\hat{p})\,, 
\end{equation}
which we want to match asymptotically onto the known solution
\begin{equation}\label{appexact}
\phi_{p'}(x) = \frac{4\pi\,\e^{\im\,E' t}}{r} \sum_{\ell, m} Y_{\ell}^{m}(\hat{x})\, \overline{Y_{\ell}^{m}}(\hat{p}\,^{\prime})\, R_{\ell m}(p';r),
\end{equation}
where we have explicitly denoted the dependence of the radial wavefunction on the momentum $p'$.

Since the spherical harmonics are orthogonal, we can equate coefficients between \eqref{ans2} and \eqref{appexact} at large distances to find 
\begin{equation}\label{appmatch1}
\lim_{r\to\infty} \int \d \Phi(p)\, \Lambda(p)\, \e^{\im\, Et}\, \sum_{\ell,m}c_\ell(r)\, Y^{m}_{\ell}(\hat{x})\,\overline{Y^{m}_{\ell}}(\hat{p})\, \frac{2\,e^{\im\,E't}}{r}\,\sum_{\ell,m}Y^{m}_{\ell}(\hat{x})\,\overline{Y^{m}_{\ell}}(\hat{p})\,R_{\ell m}(p';r)\,.
\end{equation}
This equality requires $E=E'$, and this in turn implies that $|\vec{p}| = |\vec{p}'|$. Exploiting this and the underlying spherical symmetry, \eqref{appmatch1} implies that
\be\label{lambdaexp1}
\Lambda^{p'}(p)=\delta\!\left(|\vec{p}|-|\vec{p}\,^{\prime}|\right)\sum_{\ell',m'}Y^{m'}_{\ell'}(\hat{p})\,\overline{Y^{m'}_{\ell'}}(\hat{p}\,^{\prime})\,d_{\ell' m'}(p')\,,
\ee
in terms of some as-yet-undetermined coefficients $d_{\ell' m'}(p')$. Substituting this back into \eqref{ans2}, the two mode sums are identified due to the orthogonality of the spherical harmonics in $\hat{p}$. This means that the asymptotic matching condition is reduced to:
\be\label{appmatch2}
\lim_{r\to\infty}d_{\ell m}(p')\,c_{\ell}(r) = \frac{2\,E}{|\vec{p}|^2}\,\lim_{r\to\infty}\frac{R_{\ell m}(p';r)}{r}\,,
\ee
with the factors of $E/|\vec{p}|^2$ arising from the on-shell integrals on the left-hand-side of \eqref{appmatch1}. This in turn gives the expression
\be\label{lambdaexp2}
\Lambda^{p'}(p)=\frac{2\,E}{|\vec{p}|^2}\,\delta\!\left(|\vec{p}|-|\vec{p}\,^{\prime}|\right)\sum_{\ell,m}Y^{m}_{\ell}(\hat{p})\,\overline{Y^{m}_{\ell}}(\hat{p}\,^{\prime})\,\lim_{r\to\infty}\frac{R_{\ell m}(p';r)}{r\,c_{\ell}(r)}\,,
\ee
for the matching coefficients.

To further process this expression, we need to investigate the asymptotic properties of the $c_{\ell}(r)$ defined by \eqref{legTransf}. Observe that
\be\label{cexp}
\begin{split}
c_{\ell}(r)&=\int_{0}^{\pi}\d\theta\,\sin\theta\,\e^{\im\left(-\vec{p}\cdot\vec{x}+C_p\,\log(|\vec{p}|\,r)+C_{p}\,\log(1+\cos\theta)\right)}\, P_{\ell}(\cos\theta) +O(r^{-2}) \\
 & = \e^{\im\,C_p\,\log(|\vec{p}|\,r)}\,\int_{-1}^{1}\d x\,\e^{\im\left(C_p\,\log(1+x)-|\vec{p}|\,r\,x\right)}\,P_{\ell}(x)+O(r^{-2})\,,
\end{split}
\ee
where $C_p$ is the theory-dependent constant defined by \eqref{S1em} and \eqref{S1gr}, and the integration variable is $x\equiv\cos\theta$ in the last line. Defining
\be\label{ctilde}
\tilde{c}_{\ell}(r):=\int_{-1}^{1}\d x\,\e^{\im\left(C_p\,\log(1+x)-|\vec{p}|\,r\,x\right)}\,P_{\ell}(x)\,,
\ee
we observe that
\be\label{ctildeprime}
\tilde{c}'_{\ell}(r)=-\im\,|\vec{p}|\int_{-1}^{1}\d x\,\e^{\im\left(C_p\,\log(1+x)-|\vec{p}|\,r\,x\right)}\,x\,P_{\ell}(x)\,.
\ee
Now, the recurrence relation 
\be\label{legendrecur}
(2\ell + 1)\, x\, P_{\ell}(x) = (\ell + 1)\, P_{\ell + 1}(x) + \ell\, P_{\ell - 1}(x)\,,
\ee
for Legendre polynomials can be combined with \eqref{ctilde} and \eqref{ctildeprime} to deduce the recurrence relation
\be\label{ctrecur}
\tilde{c}_{\ell+1}(r)=\frac{\im\,(2\ell+1)}{|\vec{p}|\,(\ell+1)}\,\tilde{c}'_{\ell}(r)-\frac{\ell}{\ell+1}\,\tilde{c}_{\ell-1}(r)\,,
\ee
for the $\tilde{c}_\ell$s. By computing the first few of these coefficients explicitly, one soon arrives at
\be\label{ctsol}
\tilde{c}_{\ell}(r)=\frac{\im\,\e^{\im\left(C_p\,\log 2-|\vec{p}|\,r\right)}}{|\vec{p}|\,r}+(-1)^{\ell}\,\frac{C_p\,\Gamma(\im\,C_p)\,\e^{\im\left(|\vec{p}|\,r-C_p\,\log(\im|\vec{p}|\,r)\right)}}{|\vec{p}|\,r}+O(r^{-2})\,,
\ee
which can easily be proven to solve \eqref{ctrecur} for all $\ell$ by induction. Note that for $C_p\to0$, this reproduces the asymptotic expansion for spherical Bessel functions, as expected.

In \eqref{ctsol}, we only want contributions which will contribute to the wavefunction as waves travelling like $\e^{\im(Et-|\vec{p}|\,r)}$, so we discard the second term. Note that this un-wanted term is perfectly finite; it simply has the wrong scattering behaviour. Plugging the first term of \eqref{ctsol} into \eqref{cexp} and then into \eqref{lambdaexp2} gives:
\be\label{lambdafinal}
\Lambda^{p'}(p)=-\frac{2\im\,E}{|\vec{p}|}\,\delta\!\left(|\vec{p}|-|\vec{p}\,^{\prime}|\right)\sum_{\ell,m}Y^{m}_{\ell}(\hat{p})\,\overline{Y^{m}_{\ell}}(\hat{p}\,^{\prime})\,\lim_{r\to\infty}R_{\ell m}(p';r)\,\e^{\im\left(|\vec{p}|\,r-C_p\,\log(2|\vec{p}|\,r)\right)}\,,
\ee
matching \eqref{Lambda1} from the text.

%%%%%%%%%%%%%%%%%%%%%%%%%%%%%
%%%%%%%%%%%%%%%%%%%%%%%%%%%%%
\section{The probe and classical limits of 5-point amplitudes}
\label{app:5point}
%%%%%%%%%%%%%%%%%%%%%%%%%%%%%
%
In this appendix we describe the probe and classical limits of the 5-point amplitudes in scalar QED and gravitationally coupled scalars which appear from the classical, weak field limits of our semiclassical 3-point amplitudes on Coulomb and Schwarzschild backgrounds. We will first consider the probe limit and show how this is equivalent to a background field calculation. We then take a further, classical limit, which provides the check on the semiclassical calculations of Sections~\ref{sec:photons} and~\ref{sec:gravitons}.

We begin with the scalar QED calculation. In vacuum, consider the emission of a photon, momentum $k$ and helicity $\eta$, in the scattering of two charges, one of mass $m$, momentum $p\to p'$, the other of mass $M$, momentum $P\to P'$. We can assume without loss of generality that $P_\mu = M U_\mu$ as in (\ref{CoulSchw}). The $5$-point amplitude $\cM_5$ is easily calculated in scalar QED, with the result
\be\label{eq:QED-full-5}
\begin{split}
    \cM_5 = \im & e^2 Q\, {\hat\delta}^4(P'+p'+k-p-P)\, \times \\
    &\frac{(P'+P)^\mu}{(P-P')^2}\,
    \bigg[
    -2\varepsilon_\mu
    - \frac{\varepsilon\cdot p'}{k\cdot p'}\, (p'+p+k)_\mu
    + \frac{\varepsilon\cdot p}{k\cdot p}\, (p'+p-k)_\mu \bigg] \\
    & + (p\leftrightarrow P, p'\leftrightarrow P')  \;.
\end{split}
\ee
Consider the momentum-conserving delta functions. To see how the probe limit arises, we split these into temporal and spatial pieces:
    \be
        {\delta}^3\!\left({\vec P}' +{\vec \q}\,\right)\, \delta\!\left(\sqrt{M^2+{\vec \q}^{\,2}} +k_0 -p_0 - M\right) \;, \qquad {\vec \q}:=  {\vec p}^{\,\prime} + {\vec k} -{\vec p}\;.
    \ee
We take $M$ to be large, and assume the recoil of that particle is negligible, which quantitatively means assuming the \emph{total momentum transfer} $\vec \q$ obeys ${\vec \q}^{\,\,2} \ll M^2$. This allows us to expand the square root in the temporal delta function. We similarly expand in powers of ${\vec \q}$ in the full QED amplitude. The leading order terms are those shown explicitly in (\ref{eq:QED-full-5}), which go as $1/{\vec\q}^{\,2}$. The result is, writing $\ell_\mu = k_\mu + p'_\mu -p_\mu$,
\be\label{eq:QED-probe-limit}
\begin{split}
    \mathcal{M}_5 &\to  2M\, {\hat {\delta}}^3\!\left({\vec P}^{\,\prime} +{\vec \q}\,\right) \widetilde{\mathcal{M}}_5 \;, \\
    \widetilde{\mathcal{M}}_5 &:=
    {\hat \delta}\big(U\cdot\q\big)\,
    \frac{2\im e^2 Q}{\q^2}\,
    \bigg[
    - \varepsilon\cdot U
    + \frac{\varepsilon\cdot p'}{k\cdot p'}\, U\cdot p
    - \frac{\varepsilon\cdot p}{k\cdot p}\, U\cdot p^{\prime}
    \bigg] \;.
\end{split}
\ee
It can be checked by direct calculation that $\widetilde{\mathcal{M}}_5$ is precisely the \emph{three}-point amplitude ($p\to p'$ with emission of photon $k$) on a Coulomb background, calculated to leading order in the background charge $Q$. The differences between $\mathcal{M}_5$ and $\widetilde{\mathcal{M}}_5$ are resolved at the level of physical predictions: the additional ${\hat\delta}^3$ and factor of $2M$ in $\mathcal{M}_5$ relative to $\widetilde{\mathcal{M}}_5$ are absorbed, in \emph{observables}, by state normalisation and final state integrals over the heavy particle momentum, see also~\cite{Adamo:2021rfq}. Hence (\ref{eq:QED-probe-limit}) is indeed the probe limit, physical predictions from which agree exactly with those obtained in a direct background field calculation.

In our semiclassical approach used in the text, we work to leading nontrivial order in $\hbar$. To compare to the perturabtive QED results here, we thus need to identify the leading classical behaviour of (\ref{eq:QED-probe-limit}). To do so we recall that massless momenta $k$ scale as $\hbar$ times classical wavenumber, while continuing to impose that the recoil of the mass $M$ particle is small compared to its own rest mass. Hence, we write $p'_\mu=p_\mu + \Q_\mu$ in which $\Q$ also scales as $\hbar$ to leading order~\cite{Kosower:2018adc}. Taking the classical limit of (\ref{eq:QED-probe-limit}) is then equivalent to taking a Taylor expansion in which both $k$ and $q$ are of the same small order. This expansion yields the leading classical behaviour
 \be\label{eq:QED-probe+classical}
    \widetilde{\mathcal{M}}_5 \to
    {\hat \delta}\big(U\cdot(\Q+k)\big)\,
    \frac{2\im e^2 Q}{(\Q+k)^2}
    \bigg[
    - \varepsilon\cdot U
    + \frac{U\cdot p\, \Q\cdot \varepsilon}{k\cdot p} 
    - \frac{U\cdot \Q\, p\cdot \varepsilon}{k\cdot p} 
    - \frac{k\cdot \Q\, U\cdot p\, p\cdot \varepsilon}{(k\cdot p)^2}
    \bigg] \;.
\ee
We confirm in the text that this is indeed what is recovered from our WKB analysis. 

\medskip

For gravity, the implementation of the probe and classical limits happens in exactly the same fashion; the only distinction is the functional form of the 5-point amplitude for graviton emission from the scattering of two massive scalars. Needless to say, this computation, in Feynman diagrams, is significantly more complicated than its scalar QED cousin due to the appearance of the cubic graviton vertex. In this case it is easiest to avoid Feynman diagrams by employing double copy methods~\cite{Luna:2017dtq}, which also allow for the classical limit to be taken immediately. Employing the same momentum notation as before, the classical 5-point graviton emission amplitude is
\begin{multline}\label{grav5pt}
\cM_5=16\,M^2\,\kappa^3\,\hat{\delta}^{4}\!\left(P'+p'+k-p-P\right)\,\frac{\varepsilon_{\mu\nu}}{\ell^2\,(p-p')^2}\,\times \\
\Bigg[4\left(k\cdot p\,U^\mu-k\cdot U\,p^\mu\right) \left(k\cdot p\,U^\nu-k\cdot U\,p^\nu\right)+4\,p\cdot U\,\left(k\cdot p\,U^{(\mu}-k\cdot U\,p^{(\mu}\right)\,Q^{\nu)} \\
+\left((p\cdot U)^2-\frac{m^2}{2}\right)\!\left(Q^{\mu}\,Q^{\nu}-\frac{\ell^2\,(p-p')^2}{(k\cdot p)^2\,(k\cdot U)^2}\,\left(k\cdot p\,U^\mu-k\cdot U\,p^\mu\right) \left(k\cdot p\,U^\nu-k\cdot U\,p^\nu\right)\right)\Bigg]\,,
\end{multline}
where
\be
Q^{\mu}:=-2(p'-p)^{\mu}-k^{\mu}-\frac{(p-p')^2\,p^{\mu}}{k\cdot p}+\frac{\ell^2\,U^{\mu}}{k\cdot U}\,.
\ee
Applying the probe limit to this expression, we arrive at \eqref{gpert2} in Section~\ref{sec:gravitons}.

\bibliographystyle{JHEP}
\bibliography{NewBib}

\end{document}